%% file: main.tex
\documentclass[transmag]{IEEEtran}
\usepackage{latexsym}
\usepackage{graphicx}
\usepackage{cite}
\usepackage{amsfonts,amssymb,amsmath}
\usepackage{hyperref}
\def\BibTeX{{\rm B\kern-.05em{\sc i\kern-.025em b}\kern-.08em T\kern-.1667em\lower.7ex\hbox{E}\kern-.125emX}}

\ifCLASSINFOpdf

\else

\fi

\usepackage{subcaption}
\usepackage{epsfig,psfrag}
\usepackage{mathcomp}
\usepackage{xcolor}
\usepackage[normalem]{ulem}

\makeatletter
\def\endthebibliography{%
	\def\@noitemerr{\@latex@warning{Empty `thebibliography' environment}}
	\endlist
}

\makeatother

\begin{document}

\title{Channel Hardening in Massive MIMO:\\ Model Parameters and Experimental Assessment}

\author{{Sara Gunnarsson,
		Jose Flordelis,
		Liesbet Van der Perre,
		Fredrik Tufvesson} 
	\thanks{The manuscript was submitted on 2020-02-03. This work has received funding from the strategic research area ELLIIT.}
	
	\thanks{All authors are with the Department of Electrical and Information Theory, Lund University, Lund, Sweden. S. Gunnarsson and L. Van der Perre are also with the Department of Electical Engineering, KU Leuven, Ghent, Belgium. J. Flordelis is also with Research Center Lund, Sony Corporation, Lund, Sweden (e-mail: \{Sara.Gunnarsson, Jose.Flordelis, Fredrik.Tufvesson\}@eit.lth.se, Liesbet.Vanderperre@kuleuven.be).}
	\thanks{Part of the material was presented at the IEEE International Workshop on Signal Processing Advances in Wireless Communications (SPAWC), 2018, Kalamata, Greece.}
	}

\IEEEtitleabstractindextext{\begin{abstract}
Reliability is becoming increasingly important for many applications envisioned for future wireless systems. A technology that could improve reliability in these systems is massive MIMO (Multiple-Input Multiple-Output). One reason for this is a phenomenon called channel hardening, which means that as the number of antennas in the system increases, the variations of channel gain decrease in both the time- and frequency domain. Our analysis of channel hardening is based on a joint comparison of theory, measurements and simulations. Data from measurement campaigns including both indoor and outdoor scenarios, as well as cylindrical and planar base station arrays, are analyzed. The simulation analysis includes a comparison with the COST 2100 channel model with its massive MIMO extension. The conclusion is that the COST 2100 model is well suited to represent real scenarios, and provides a reasonable match to actual measurements up to the uncertainty of antenna patterns and user interaction. Also, the channel hardening effect in practical massive MIMO channels is less pronounced than in complex independent and identically distributed (i.i.d.) Gaussian channels, which are often considered in theoretical work.
\end{abstract}
\begin{IEEEkeywords}
Channel hardening, channel model, COST 2100 channel model, massive MIMO, measurements, reliability
\end{IEEEkeywords}}

\maketitle

\section{Introduction}
\input{introduction.tex}

\section{Measurement scenarios and equipment}
\input{measurements.tex}

\section{Channel hardening in theory}
\input{channelhardening.tex}

\section{Channel hardening in practice}
\input{importantparameters.tex}

\section{Channel hardening in the COST 2100 channel model}
\input{cost2100.tex}

\section{Comparison between theory, measurements and simulations}
\input{comparison.tex}

\section{Conclusion}
\input{conclusion.tex}

\bibliographystyle{IEEEtran}

\bibliography{my_bib}

\end{document}

%% file: introduction.tex
\IEEEPARstart{M}{ission-critical} applications such as remote surgery, intelligent transportation systems and industry automation are envisioned to be based on wireless connectivity in the future. To realize these applications, reliable communication is required. Massive MIMO \cite{Marzetta2010}\cite{mamimo} has in theory shown a potential to increase reliability; by deploying a massive number of antennas at the base station side, an unprecedented level of spatial diversity can be exploited. Additional advantages are that both spectral and energy efficiency increase. However, the enhanced reliability is the main focus of this paper.

Stable channels are essential in order to achieve reliability in communication systems. In massive MIMO systems, the channel gain becomes more concentrated around its mean when increasing the number of base station antennas. This phenomenon, where a fading channel behaves more deterministically, is called channel hardening. The decreasing fading variations will also render a more stable capacity. The channel hardening effect can be studied from two points of view. The first point is that the fading over frequency is reduced due to the decrease of experienced delay spread. The second point is channel hardening in the time domain, where the temporal fading decreases as a result of the coherent combination of the signals from the many base station antennas.

The theory of channel hardening has been discussed in many papers, see e.g. \cite{Marzetta2010, Bjoernson2017,downlink_pilots, chhard_phy_model,8902768,proof_conv}. In \cite{Bjoernson2017} and \cite{downlink_pilots} a definition of channel hardening is given and the authors in \cite{chhard_phy_model} introduce and derive closed form results for the coefficient of variation of channel gain, which further relates to the characteristics of the channel, which are affecting channel hardening. In \cite{proof_conv}, a proof of complete convergence of channel hardening in massive MIMO with uncorrelated Rayleigh fading is presented. Although co-located massive MIMO systems are shown to provide channel hardening, this might not be the case in cell-free massive MIMO \cite{cell_free}. Much research has been carried out on the theoretical aspects of channel hardening, not as many studies have investigated this phenomenon experimentally \cite{chhard_spawc, delay_spread_mamimo, chhard_aalborg, chhard_aalborg_2, 8581026, Ghiaasi2019}.

Early theoretical massive MIMO studies and proofs of channel hardening have been relying on the assumption that the channels experience such rich scattering that they can be modeled as complex independent and identically distributed (i.i.d.) Gaussian channels. However, it has been shown in measurements that this is not the case \cite{meas_xiang}\cite{ant_equal} in practice. In general, real massive MIMO channels are spatially correlated. This phenomenon has been acknowledged and recently it has also been taken into account in theoretical analyses of channel hardening \cite{Bjoernson2017}\cite{mamimo_20}. The other extreme channel is the keyhole channel, which is shown to not provide any channel hardening \cite{downlink_pilots}. Any real massive MIMO system will experience a channel hardening somewhere in between; here we investigate what actually can be expected in different real scenarios.

To develop and assess the system performance of future mission-critical applications, adequate channel models that can capture the relevant channel characteristics are needed. However, the relation between experiments and models has not been studied so far. Several channel models could be used for this purpose, such as the QuaDRiGa channel model \cite{quadriga}\cite{quadriga_2} with extensions for WINNER-type models \cite{winner}, the 3GPP channel model (TR 38.901) \cite{3gpp_chmodel}, the IMT-2020 channel model \cite{itu}, and the METIS channel model \cite{metis}. Here, the COST 2100 channel model \cite{cost2100}\cite{cost_model} is chosen as it is a geometry-based stochastic channel model (GSCM) with the important property that it can consistently describe the channel in space, time and frequency. It also realistically captures the behavior at a multipath level. The model has recently been extended with features such that it realistically can reflect characteristics that are prominent in measured massive MIMO channels \cite{cost_extension}\cite{cost_pepe}. To serve as an adequate channel model for the intended applications, all important channel aspects that affect system performance should be caught by the model; this leads to the questions if and how the channel model realistically can capture the channel hardening effect as well, and what are appropriate model parameters.

The contribution of this paper is twofold. First, channel hardening has been validated by experiments; this being a key part in the process of extracting relevant parameters that can be used for accurate channel modeling. We extend our analysis in \cite{chhard_spawc} with evaluation of more parameters and thereby providing further insights; the number of analyzed scenarios has been extended and the analysis is now also including outdoor environments as well as different antenna arrays. We present generalized values for what can be expected in terms of channel hardening in a co-located massive MIMO system. The second contribution, novel in relation to \cite{chhard_spawc}, is an assessment of the capabilities of the COST 2100 channel model with its massive MIMO extension to capture the channel hardening effect and an evaluation of its model parameters. This is evaluated by relating simulated results from the model to our experimental findings. We highlight key parameters, related to propagation and antenna characteristics, that affect the channel hardening. With this knowledge, appropriate channel models and parameter sets can be selected and used to acquire realistic massive MIMO channels; consequently, these could be appropriate for development and assessment of future wireless applications.

The structure of the paper is as follows; Section~II describes the measurement campaigns, including both the scenarios and the equipment used. Section~III presents theory and the definition of channel hardening. The measurement results, including generalized values for channel hardening in different scenarios, are presented in Section~IV. Section~V describes the COST 2100 channel model and the performed simulations. Section~VI presents a joint comparison between theory, measurements and simulations in order to give a more thorough and complete analysis of channel hardening in co-located massive MIMO systems. In Section~VII, the paper is summarized and conclusions are presented.

%% file: measurements.tex
\begin{figure}[t]
	\centering
    \includegraphics[width=3in]{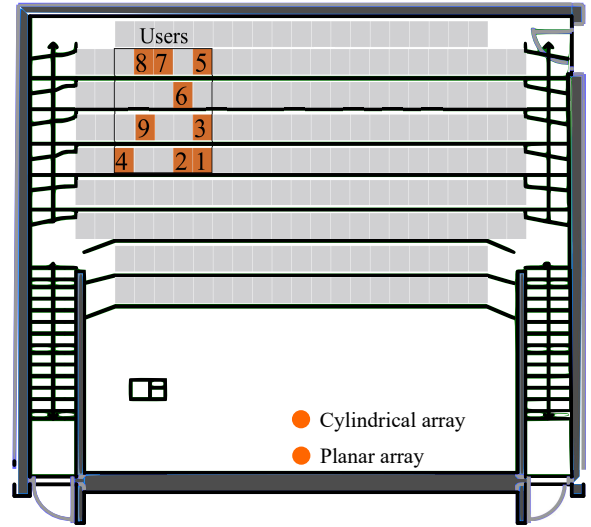}
	\caption{The auditorium where the indoor scenario took place. The base station is standing in the front of the room, at different positions depending on the array used in the measurements, and the users are sitting in the back of the room to the left.}
	\label{fig:location}
\end{figure}

\begin{figure}[t]
	\centering
	\includegraphics[width=3in]{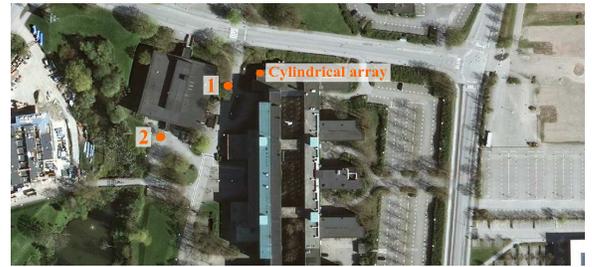}
	\caption{The outdoor scenario for both LOS (position 1) and NLOS (position 2). The base station has a cylindrical array and is positioned on the roof. The users were moving within a circle with a diameter of 5~m.}
	\label{fig:location2}
\end{figure}

\begin{figure}[t]
	\centering
	\includegraphics[width=3in]{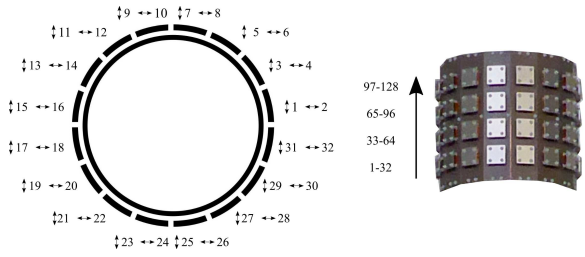}
	\caption{The cylindrical base station antenna array as seen from above (left) with the numbering of the antenna elements in the first ring, both vertically and horizontally polarized. The cylindrical array seen from the side (right) with the numbering per ring.}
	\label{fig:cyl_array}
\end{figure}

\begin{figure}[t]
	\centering
    \includegraphics[width=3in]{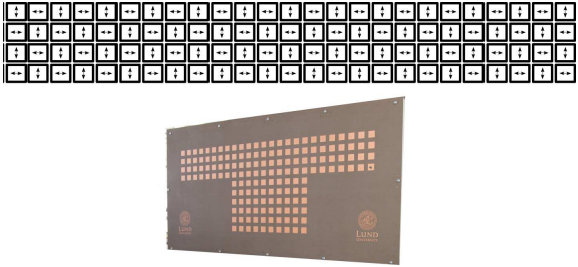}
	\caption{The planar base station antenna array as seen from the front (lower). Every other antenna is vertically and horizontally polarized (upper), where the 100~connected antennas are shown with their corresponding polarizations.}
	\label{fig:lumami}
\end{figure}

The main focus of this work is on indoor propagation, as this typically is the scenario with the richest scattering, and therefore the environment where channel hardening is most pronounced. Only single-antenna user equipment is considered, limiting this study to focus on how channel hardening is affected when only changing the number of base station antennas. For this primary scenario, the base station is situated in the front of an indoor auditorium, and is serving a group of nine closely-spaced users. We also consider another scenario, which is an outdoor scenario where a base station is located on a roof serving nine closely-spaced users. Two different measurement campaigns have been conducted and the measured channels have been analyzed in terms of channel hardening. The first measurement campaign considered both scenarios and the measurements were performed with a channel sounder deployed with a cylindrical array. This campaign and the performed measurements therein is further described in \cite{Bourdoux2015}. The second measurement campaign aimed at repeating and validating some of the measurements in the indoor scenario with a real-time testbed equipped with a planar array. For all measurements the user equipment antenna was a vertically polarized SkyCross SMT-2TO6MB-A.

\subsection{Indoor scenario}
The indoor measurement campaign was carried out in an auditorium located at Lund University. As seen in Fig.~\ref{fig:location}, a base station is standing in the front of the room and is serving users sitting in the left back corner. With the cylindrical array, simultaneous measurements were done when serving all nine users. With the planar array, evaluations were performed for user~1 and 5, which serve as representatives of users in the front and in the back rows, respectively. The users were mostly static during the measurements but were moving the antennas slowly back and forth, tilted approximately 45~degrees, with a speed lower than 0.5~m/s. The measurement campaigns also included, for both antenna arrays, the case when the rectangle in Fig.~\ref{fig:location} was almost full with people in the vicinity of the active users.

\subsection{Outdoor scenario}
In the outdoor scenario, the measurement campaign was carried out in an open area at the campus of Lund University. A base station was placed at the roof on top of the second floor and measurements were performed both for a Line-Of-Sight (LOS) scenario (position~1 in Fig.~\ref{fig:location2}) and for a non-LOS (NLOS) scenario (position~2 in Fig.~\ref{fig:location2}). In both cases, nine closely-spaced users were moving around within a circle having a diameter of 5~m. The users were holding the antennas tilted approximately 45~degrees.

\subsection{Channel sounder}
The RUSK LUND MIMO channel sounder was used for the first measurement campaign with the cylindrical array. This channel sounder is a multiplexed-array channel sounder, meaning that the transfer functions for each transmit-receive antenna pair is measured at a fast pace one after the other. The cylindrical array connected to the channel sounder is shown in Fig.~\ref{fig:cyl_array}. The array consists of 64~dual-polarized patch antennas spaced half a wavelength apart at the carrier frequency of 2.6~GHz, resulting in 128~antenna ports in total. The bandwidth is 40~MHz. The antenna elements are in the measurement data numbered according to Fig.~\ref{fig:cyl_array}, i.e. starting with the lower ring and finishing with the upper ring as seen to the right in the figure. To the left in Fig.~\ref{fig:cyl_array}, the lower ring is seen as from above. Important to note is that odd-numbered antenna ports are vertically polarized, even-numbered ones have a horizontal polarization. For the indoor scenario the measurements produced 129~frequency points and 300~snapshots taken over 17~seconds. This was also done for the outdoor scenario, with the exception that here 257~frequency points were measured. For the indoor scenario, the time for sounding is 3.2~$\mu$s for one transmit-receive antenna pair, leading to a time per snapshot of approximately 7.37~ms. For the outdoor scenario, these values doubles due to measuring twice as many frequency points.

\subsection{Real-time testbed}
The Lund University Massive MIMO testbed (LuMaMi) was used for the second measurement campaign with the planar array. This testbed is based on software-defined radio technology and is operating in real-time at a carrier frequency of 3.7~GHz with 20~MHz of bandwidth. More information about the technical details of the testbed can be found in \cite{lumami} and \cite{lumami_journal}. The planar array connected to the testbed is shown in Fig.~\ref{fig:lumami}, where the lower part shows the array from the front. The upper four rows with 25~antennas at each row are connected to one transceiver chain each, resulting in 100~antenna elements, spaced half a wavelength apart. Every other antenna element is vertically polarized while the other element has a horizontal polarization. The measurements when using the testbed produced 100~frequency points per user and for 20~seconds, 2000~snapshots were stored\footnote{For completeness, it should be noted that 9 snapshots were lost during the measurements with user~1, and 7 snapshots during the measurements with user~5, possibly due to bad synchronization. These snapshots were therefore removed when analysing the data.}.

%% file: channelhardening.tex
Following \cite{downlink_pilots}, let $\mathbf{h}_{k}$ be the channel vector between the base station and user $k$. The channel $\mathbf{h}_{k}$ offers hardening if

\begin{equation}
\frac{\text{Var}\{\|\mathbf{h}_{k}\|^2\}}{\text{E}\{\|\mathbf{h}_{k}\|^2\}^2}\rightarrow 0, \hspace{0.5cm} \text{as} \hspace{0.2cm} M\rightarrow \infty,
\label{eq:chhard_def}
\end{equation}

\noindent where $M$ is the number of base station antennas. In \cite{chhard_phy_model}, the coefficient of variation ($CV^2$) is introduced and further derived as

\begin{equation}
\begin{split}
&CV^2=\frac{\text{Var}\{\|\mathbf{H}\|^2_F\}}{\text{E}\{\|\mathbf{H}\|^2_F\}^2} \\ =
&\mathcal{E}^2(\mathcal{A}_{tx},\mathcal{D}_{tx})\mathcal{E}^2(\mathcal{A}_{rx},\mathcal{D}_{rx})\frac{\text{E}\{\|\mathbf{c}\|^4-\|\mathbf{c}\|^4_4\}}{\text{E}\{\|\mathbf{c}\|^2\}^2}
+\frac{\text{Var}\{\|\mathbf{c}\|^2\}}{\text{E}\{\|\mathbf{c}\|^2\}^2},
\end{split}
\label{eq:coeff_var}
\end{equation}

\begin{figure}[b]
	\centering

    \psfrag{Number of base station antennas}[][][1.2]{Number of base station antennas}
    \psfrag{Standard deviation [dB]}[][][1.2]{Standard deviation [dB]}
    \psfrag{Number of physical paths}[][][1.2]{\hspace{2.5cm}Number of physical paths}
    \resizebox{3in}{!}{\includegraphics[width=4in]{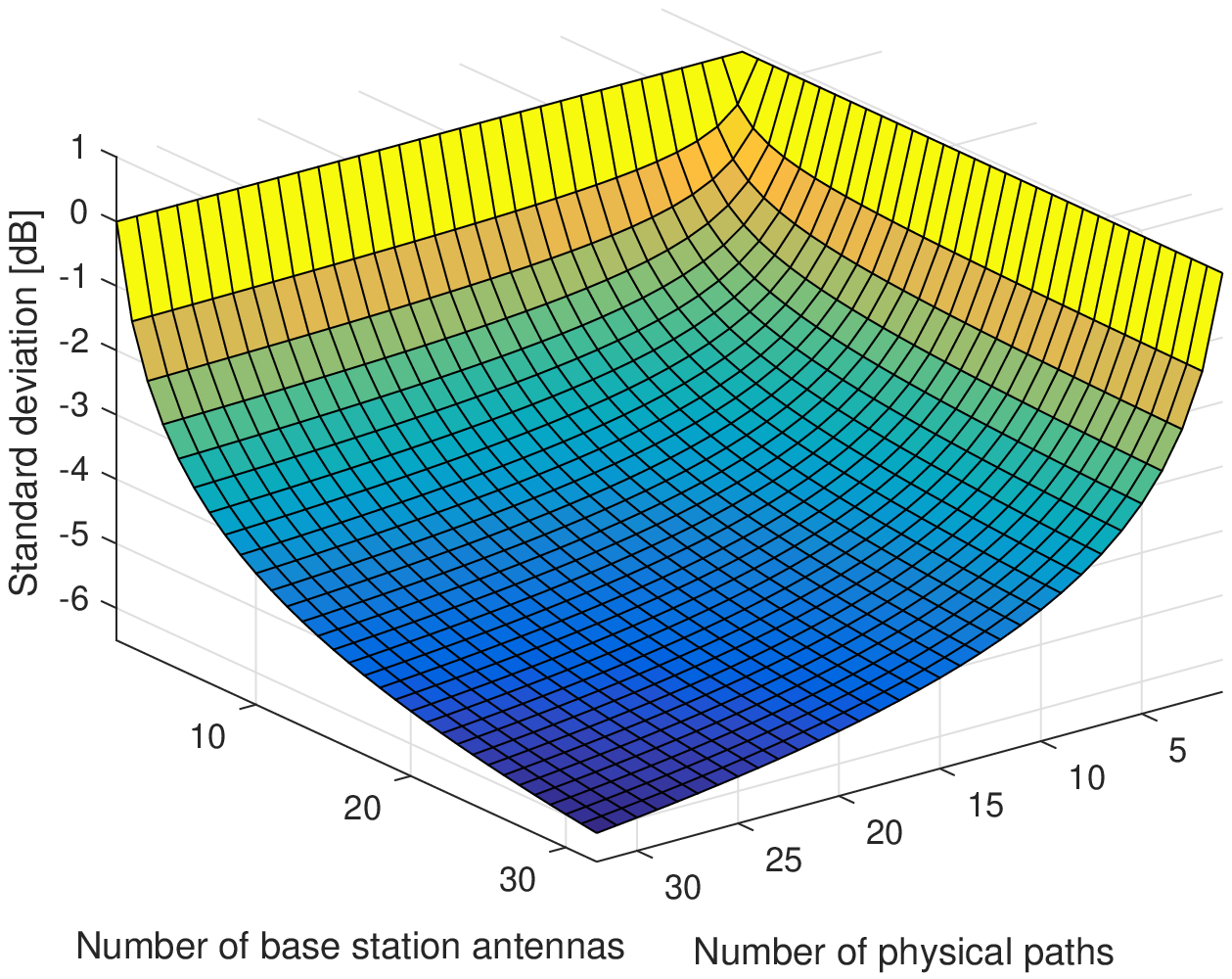}}
	\caption{Standard deviation of channel gain as a function of number of base station antennas and the number of physical paths. Interpretation based on \cite{chhard_phy_model}.}
	\label{fig:tradeoff}
\end{figure}

\noindent where $\|\mathbf{H}\|_F$ is the Frobenius norm of $\mathbf{H}$, $\mathcal{E}^2(\mathcal{A}_{tx},\mathcal{D}_{tx})$ and $\mathcal{E}^2(\mathcal{A}_{rx},\mathcal{D}_{rx})$ are the second moments of the inner products between the steering vectors of two different physical paths for the transmitter and receiver, respectively, and $\|\mathbf{c}\|^2$ is the accumulated channel gain from all physical paths. In our case $\mathbf{H}=\mathbf{h}_k$. The first term in (\ref{eq:coeff_var}) can be considered as a small-scale fading factor. The second term can be considered as a large-scale fading factor, which for channel coefficients with independent channel gains with variance $\sigma^2$ and mean $\mu$ is shown to be equal to $\frac{1}{P}(\frac{\sigma}{\mu})^2$ \cite{chhard_phy_model}, where~$P$ is the number of physical paths in the environment. Under the assumptions that the channel coefficients for all physical paths are $\mathcal{CN}(0,1)$ and that the steering vectors of the physical paths are distributed uniformly over the unit sphere, (\ref{eq:coeff_var}) can be further rewritten as

\begin{equation}
CV^2 = \frac{1}{M}(1-1/P)+1/P
\label{eq:cv}
\end{equation}

\noindent where we here consider $M$ base station antennas and $1$ user antenna \cite{chhard_phy_model}. The standard deviation of the channel gain as a function of the number of base station antennas and physical paths is shown in Fig.~\ref{fig:tradeoff}, obtained through a simulation based on (\ref{eq:cv}) and then taking the square-root of the results. As expected, more base station antennas increase the channel hardening, although this is only true if there are enough physical paths in the environment. The achievable channel hardening in a specific environment will therefore be limited either by the number of base station antennas or the number of physical paths, assuming a fixed number of antennas for the user equipment. 
It is worth noting in (\ref{eq:cv}) that if the number of physical paths $P\rightarrow \infty$, then the coefficient of variation $CV^2\rightarrow \frac{1}{M}$. If also the number of antennas $M\rightarrow \infty$ then $CV^2\rightarrow 0$, and the channel offers hardening as defined in (\ref{eq:chhard_def}).

The data, acquired as previously described for the two measurement campaigns, has been analyzed in order to validate the theoretical channel hardening results. This is done in the same way as in \cite{chhard_spawc}, and similar to the investigation in \cite{chhard_aalborg}\cite{chhard_aalborg_2}. Specifically, the measured channel transfer functions have been normalized according to

\begin{equation}
\mathbf{\overline{h}}_{k}(n,f) = \frac{\mathbf{h}_{k}(n,f)}{\sqrt[]{\frac{1}{N F M}\sum_{n=1}^{N}\sum_{f=1}^F\sum_{m=1}^M|h_{km}(n,f)|^2}},
\label{eq:norm}
\end{equation}

\noindent where $N$ is the number of snapshots, $F$ is the number of frequency points and $M$ is the number of \textit{selected} base station antennas, i.e. those over which the standard deviation is later computed in order to obtain a measure of the experienced channel hardening. This normalization makes sure that the average power of each entry in $\overline{\mathbf{h}}_{k}$, averaged over frequency, time, and base station antennas, is equal to one.

For $M$ selected base station antennas, the instantaneous channel gain for each user is defined as

\begin{equation}
\overline{G}_k(n,f) = \frac{1}{M}\sum_{m=1}^{M} |\overline{h}_{km}(n,f)|^2,
\label{eq:subset_pwr}
\end{equation}

\noindent such that the average channel gain

\begin{equation}
\mu_k = \frac{1}{N F} \sum_{n=1}^N \sum_{f=1}^F \overline{G}_k(n,f) = 1
\label{eq:mean}
\end{equation}

\noindent is independent of the number of antennas selected at the base station. This means that the base station can reduce the total output power with a factor of $M$, i.e. the beamforming gain. The standard deviation of channel gain is computed for each user according to

\begin{equation}
\text{std}_k = \sqrt{\frac{1}{N F} \sum_{n=1}^{N} \sum_{f=1}^{F} |\overline{G}_k(n,f) - \mu_k|^2},
\label{eq:std}
\end{equation}

\noindent where the instantaneous channel gain for user $k$, $\overline{G}_k(n,f)$, is given in (\ref{eq:subset_pwr}) and the average channel gain for user $k$, $\mu_k$, is given in (\ref{eq:mean}). The standard deviation in (\ref{eq:std}) is indeed an estimate, $\hat{CV}(M)$, of the (square root of the) coefficient of variation (\ref{eq:cv}), for some $M\ge1$. In the following, when quantifying the channel hardening for some subset of antennas of size $M$, we use the difference

\begin{equation}
    \hat{CV}(M)-\hat{CV}(1)
    \label{eq:chhard}
\end{equation}

\noindent of the standard deviation as given in (\ref{eq:std}). This means that depending on which antenna element that is chosen as the reference element, the channel hardening will result in different values. How this reference element is chosen in our analysis will be elaborated on in the next section.

%% file: importantparameters.tex
This section presents our contribution to the experimental research on channel hardening in real environments, extending the initial results in \cite{chhard_spawc} to include more scenarios, array geometries and propagation characteristics. In addition, we extend the analysis and provide further insights on aspects that impact the experienced channel hardening and hence, the overall reliability in massive MIMO systems.

\begin{figure}[t]
	\centering
    \psfrag{20}[][]{\small 20}
    \psfrag{60}[][]{\small 60}
    \psfrag{100}[][]{\small 100}
    \psfrag{-60}[][]{\small -60}
    \psfrag{-65}[][]{\small -65}
    \psfrag{-70}[][]{\small -70}
    \psfrag{-75}[][]{\small -75}
    \psfrag{Base station antenna}[][][0.8]{\Large Base station antenna}
    \psfrag{Mean channel gain [dB]}[][][0.8]{\Large Average channel gain [dB]}
    \resizebox{3.5in}{!}{\includegraphics[width=4in]{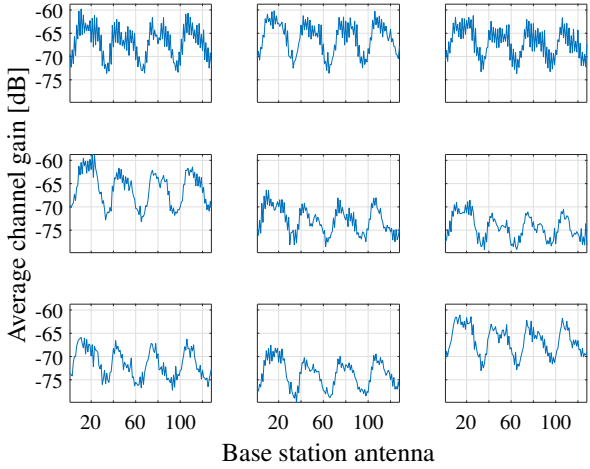}}
	\caption{Unnormalized average channel gain for the 128~base station antennas in the cylindrical array shown for all nine users. Numbering of the users is row-wise, starting from the top left corner.}
	\label{fig:mean_pwr}
\end{figure}

For the indoor scenario with the cylindrical array, see Fig.~\ref{fig:location} and Fig.~\ref{fig:cyl_array}, the un-normalized average channel gain for each of the 128~base station antennas is shown in Fig.~\ref{fig:mean_pwr}, for all nine users. Since the channels are not normalized, what is seen are the actual measured differences in average channel gain between users and antenna elements in the array. As expected, there are for each user large variations over the array which can be seen as the four larger peaks and dips, one for each ring of antennas, due to some antennas experiencing a LOS condition while some are not. Fig.~\ref{fig:cyl_array} is used as reference for the numbering of the antennas in the rings as well as the numbering between the two polarizations. The latter explains the more local variations in Fig.~\ref{fig:mean_pwr}, where the average channel gain is alternating between every two consecutive antennas. Both these variations contributes to an imbalance between the antenna elements in the array. The results in Fig.~\ref{fig:mean_pwr} can also be explained as an effect of the interaction between the cylindrical dual-polarized array and the environment. The variations that are seen show that the multipath components are not coming in with equal strength from all directions. They are rather coming dominantly from some distinct angles, in line with other experimental studies \cite{meas_xiang}\cite{ant_equal}, and the previous analysis in \cite{Bourdoux2015}, where also the distribution of channel coefficients can be found. Detailed cluster analysis can be found in \cite{cost_pepe} This observation questions the appropriateness of modeling massive MIMO channels with the complex Gaussian channel model and strengthens the conclusion that massive MIMO channels are indeed spatially correlated.

\begin{figure}[t]
	\centering
	\resizebox{3in}{!}{\includegraphics[width=4in]{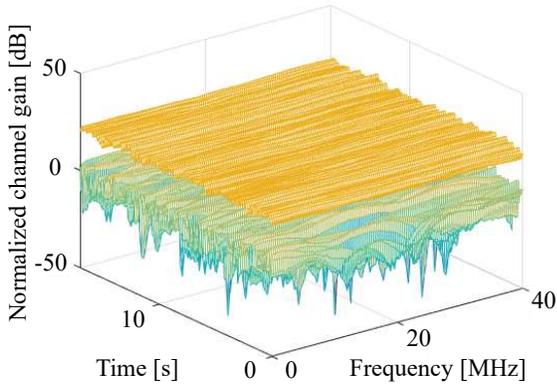}}
	\caption{Normalized channel gain for user~1 showing the single antenna with the median average channel gain (lower) and the total channel gain for all 128~base station antennas in the cylindrical array (upper).}
	\label{fig:3D_best_1}
\end{figure}

\begin{figure}[t]
	\centering
	\vspace{0.25in}
    \psfrag{2}[][]{\Large 2}
    \psfrag{0}[][]{\Large 0}
    \psfrag{-2}[][]{\Large -2}
    \psfrag{-4}[][]{\Large -4}
    \psfrag{-6}[][]{\Large -6}
    \psfrag{-8}[][]{\Large -8}
    \psfrag{-10}[][]{\Large -10}
    \psfrag{10}[][]{\Large 10 \hspace{0.025in}}
    \psfrag{1}[][]{\Large 1}
    \psfrag{Number of base station antennas}[][][1.1]{\Large Number of base station antennas}
    \psfrag{Standard deviation [dB]}[][][1.1]{\Large Standard deviation [dB]}
    \psfrag{Gaussian}[][]{\normalsize Gaussian}
    \psfrag{Original}[][]{\normalsize Original}
    \psfrag{Strongest first}[][]{\normalsize Strongest first}
    \psfrag{Weakest first}[][]{\normalsize Weakest first}
    \resizebox{3in}{!}{\includegraphics[width=4in]{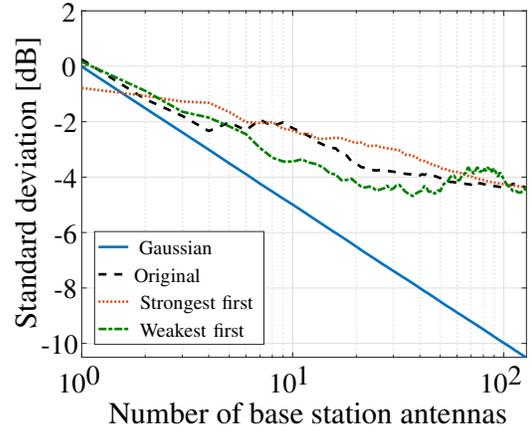}}
	\caption{Standard deviation of channel gain as a function of number of base station antennas when selecting antennas in different orders, for user~1 and the cylindrical base station array, and the complex Gaussian channel as reference.}
	\label{fig:comparison1}
\end{figure}

\subsection{Channel hardening in time and frequency}

A clear visualization of the channel hardening effect is seen in Fig.~\ref{fig:3D_best_1}, where the normalized channel gain with a single base station antenna versus the channel gain when using all 128~base station antennas is shown for user~1 in the indoor scenario. For the single antenna case, $M=1$, the antenna element in the base station array which has the median average channel gain is chosen. For this single antenna, there are large dips in both time and frequency. The combination of all 128~antennas results in both array gain, yielding a higher average, and smaller variations relative to this average. It is evident that in the time domain, there is a channel hardening effect. The channel flattens out when using a large number of base station antennas; the remaining variations are caused by the movements of the user equipment. The channel hardening is even more evident in the frequency domain; for the measured scenario and the considered bandwidth, the channel response becomes almost entirely flat. For comparison, in this scenario, the difference between the maximum and minimum normalized channel gain in the time domain, averaged over subcarriers and for $M=1$ and $M=128$ respectively, is 36~dB and 9~dB. The corresponding values for the frequency domain are 16~dB and 2~dB. In general in massive MIMO systems, the result will be similar as it is straight-forward to apply precoding to flatten out the channel in the frequency domain. However, shadowing effects in the time domain can not easily be compensated for.

\subsection{Channel hardening with different subsets of antennas}

The standard deviation of channel gain as the number of base station antennas~$M$ increases from 1 to 128 is shown in Fig.~\ref{fig:comparison1} for user~1. The channels are normalized according to (\ref{eq:norm}) and the instantaneous channel gain for every subset is computed for all frequencies and snapshots as in (\ref{eq:subset_pwr}) before the standard deviation of channel gain for each subset is computed using (\ref{eq:std}). The blue solid line shows the result for the complex Gaussian channel, which for 128~base station antennas has a channel hardening, measured as the decrease in standard deviation of channel gain when going from 1 to 128~base station antennas as in (\ref{eq:chhard}), of around $10.5$~dB. This is close to the theoretical value of $10\log_{10}(\sqrt{128})$~dB. The standard deviation of channel gain is also shown in Fig.~\ref{fig:comparison1} when selecting the antennas in different orders as the number of base station antennas increases. All the curves for the measured responses end up at the same point but they evolve differently as the number of antennas increases and have different reference elements and therefore different starting points.

The case 'Original' is where the antennas are chosen in the order shown in Fig.~\ref{fig:cyl_array}. For the indoor scenario, the first few selected antennas are in NLOS and then some antennas in LOS get included in the selected subset resulting in a temporary increase of the standard deviation, as seen for the black dashed line. The case 'Strongest first' means choosing the antennas with the highest average channel gain first, as computed for all frequency points and snapshots. This seems to be the most reasonable choice and leads to quite a steady decrease of the standard deviation in logarithmic scale. For comparison, the case 'Weakest first' is also shown, which is simply the reverse of the 'Strongest first' label. This results in a increase of the standard deviation when the antennas with higher channel gain, relative to the already included antennas, also are included in the subset. An observation from Fig.~\ref{fig:comparison1} is that whether the subset of selected antennas is in LOS, NLOS or a combination therefore, affects the behavior of the standard deviation curve, both in terms of starting point and regarding the course of the curve. For further analysis see \cite{chhard_spawc}. An additional remark is that even though the order 'Weakest first' has the least variations around its average for $M$ in the range from 5 to 50~antennas, and thus offers the most channel hardening, one should note that this still may not be the option to go for since the power level is normalized for the selected antennas. This makes them comparable in terms of variations relative to their respective means but that does not guarantee that choosing the weakest antenna first results in a good enough channel in terms of absolute channel gain. As an example, for $M=10$, the un-normalized total channel gain, as summed for the first ten antennas and the whole bandwidth, is -30~dB for 'Strongest first' and -42~dB for 'Weakest first', averaged over snapshots. Therefore, only the 'Strongest first' order will be used in the analysis from now on, in order to avoid any confusion.

\begin{figure}[t]
	\centering
    \psfrag{2}[][]{\Large 2}
    \psfrag{0}[][]{\Large 0}
    \psfrag{-2}[][]{\Large -2}
    \psfrag{-4}[][]{\Large -4}
    \psfrag{-6}[][]{\Large -6}
    \psfrag{-8}[][]{\Large -8}
    \psfrag{-10}[][]{\Large -10}
    \psfrag{10}[][]{\Large 10 \hspace{0.025in}}
    \psfrag{1}[][]{\Large 1}
    \psfrag{Number of base station antennas}[][][1.1]{\Large Number of base station antennas}
    \psfrag{Standard deviation [dB]}[][][1.1]{\Large Standard deviation [dB]}
    \psfrag{Gaussian}[][]{\normalsize Gaussian}
    \psfrag{Both polarizations}[][]{\normalsize Both polarizations}
    \psfrag{Vertical polarization}[][]{\normalsize Vertical polarization}
    \psfrag{Horizontal polarization}[][]{\normalsize Horizontal polarization}
    \resizebox{3in}{!}{\includegraphics[width=4in]{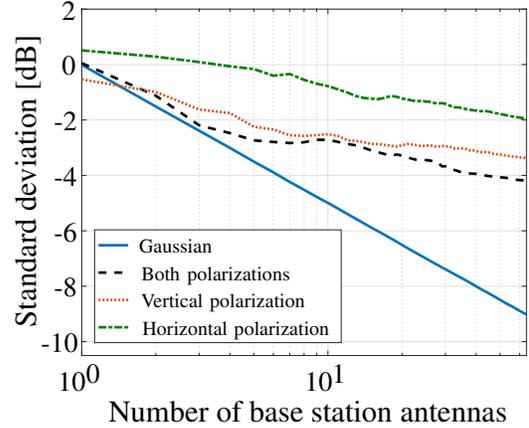}}
	\caption{Standard deviation of channel gain as a function of number of base station antennas when using one of the polarizations or a combination of both, for user~1 and the cylindrical base station array, and the complex Gaussian channel as reference.}
	\label{fig:comparison_pol}
\end{figure}

As previously indicated, the polarization state is causing variations in the channel gain over the array. Computing the standard deviation for $M=1,\ldots,64$ when using one of the polarizations or a combination of both yields the results in Fig.~\ref{fig:comparison_pol}. For 'Both polarizations', the set of antennas are selected as one polarization per antenna, starting with vertical polarization, and then alternating between the two polarizations as it traverses through the rings. This means that the three options have the same aperture. The experienced channel hardening depends on the polarization set selected, and thereby also on the selected reference element. Fig.~\ref{fig:comparison_pol} shows the decrease of standard deviation when going from 1 to 64 antennas, using the definition in (\ref{eq:chhard}). When using the 64~horizontally polarized antennas, the channel hardening is 2.7~dB, while the case with the 64~vertically polarized antennas gives a channel hardening of 2.8~dB. It is worth noting that even though the channel hardening is almost the same for these two curves, the specific value of the standard deviation of channel gain at each point is smaller when using the vertical polarization since these antennas already have smaller variations relative to its average compared to the horizontally polarized antennas. An interpretation of this is that the channel already can be considered as 'pre-hardend', which could be a result of the LOS component being mainly vertically polarized. This is supported by the fact that when computing the mean of the un-normalized average channel gain, this results in -64~dB for the vertically polarized antennas and -68~dB for the horizontally polarized antennas. %
The most channel hardening, and also the lowest standard deviation of channel gain for $M=128$, is achieved when using a combination of both polarizations; then the channel hardening is 4.2~dB. This is partly explained by the ability to exploit polarization diversity in the environment, i.e. an increased number of physical paths, but is also an effect of having different reference elements. 

\subsection{Analysis of different scenarios}

\begin{figure*}[t]
\centering
    \begin{subfigure}[b]{0.4\textwidth}
    \psfrag{2}[][]{\Large 2}
    \psfrag{0}[][]{\Large 0}
    \psfrag{-2}[][]{\Large -2}
    \psfrag{-4}[][]{\Large -4}
    \psfrag{-6}[][]{\Large -6}
    \psfrag{-8}[][]{\Large -8}
    \psfrag{-10}[][]{\Large -10}
    \psfrag{10}[][]{\Large 10 \hspace{0.025in}}
    \psfrag{1}[][]{\Large 1}
    \psfrag{Number of base station antennas}[][][1.1]{\Large Number of base station antennas}
    \psfrag{Standard deviation [dB]}[][][1.1]{\Large Standard deviation [dB]}
    \psfrag{Gaussian}[][]{\normalsize Gaussian}
    \psfrag{Indoor}[][]{\normalsize Indoor}
    \psfrag{Outdoor LOS}[][]{\normalsize Outdoor LOS}
    \psfrag{Outdoor NLOS}[][]{\normalsize Outdoor NLOS}
    \resizebox{3in}{!}{\includegraphics[width=4in]{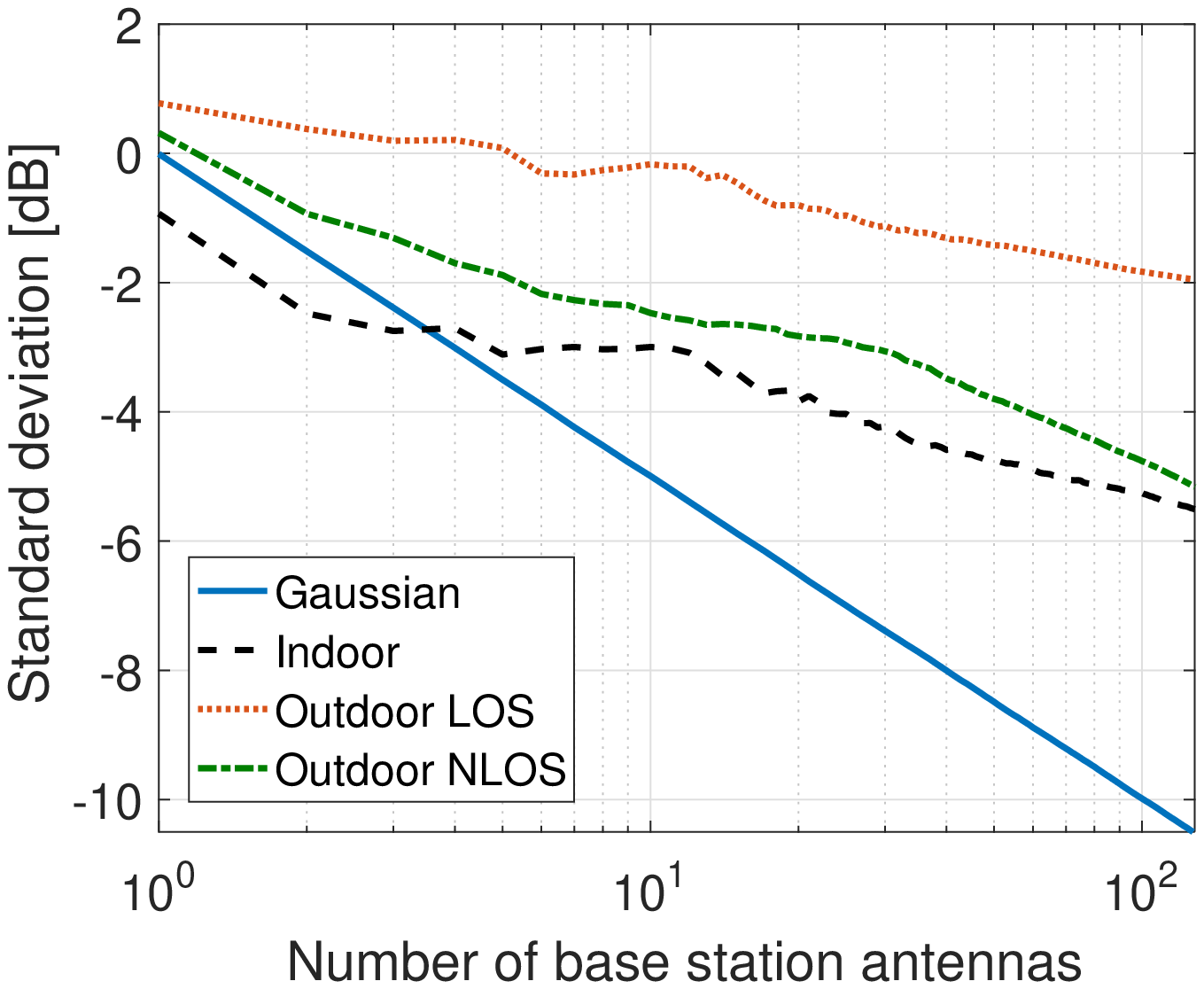}}
	\caption{The users with the most channel hardening}
	\label{fig:in_vs_out_best}
\end{subfigure}
\begin{subfigure}[b]{0.4\textwidth}
    \psfrag{2}[][]{\Large 2}
    \psfrag{0}[][]{\Large 0}
    \psfrag{-2}[][]{\Large -2}
    \psfrag{-4}[][]{\Large -4}
    \psfrag{-6}[][]{\Large -6}
    \psfrag{-8}[][]{\Large -8}
    \psfrag{-10}[][]{\Large -10}
    \psfrag{10}[][]{\Large 10 \hspace{0.025in}}
    \psfrag{1}[][]{\Large 1}
    \psfrag{Number of base station antennas}[][][1.1]{\Large Number of base station antennas}
    \psfrag{Standard deviation [dB]}[][][1.1]{\Large Standard deviation [dB]}
    \psfrag{Gaussian}[][]{\normalsize Gaussian}
    \psfrag{Indoor}[][]{\normalsize Indoor}
    \psfrag{Outdoor LOS}[][]{\normalsize Outdoor LOS}
    \psfrag{Outdoor NLOS}[][]{\normalsize Outdoor NLOS}
    \resizebox{3in}{!}{\includegraphics[width=4in]{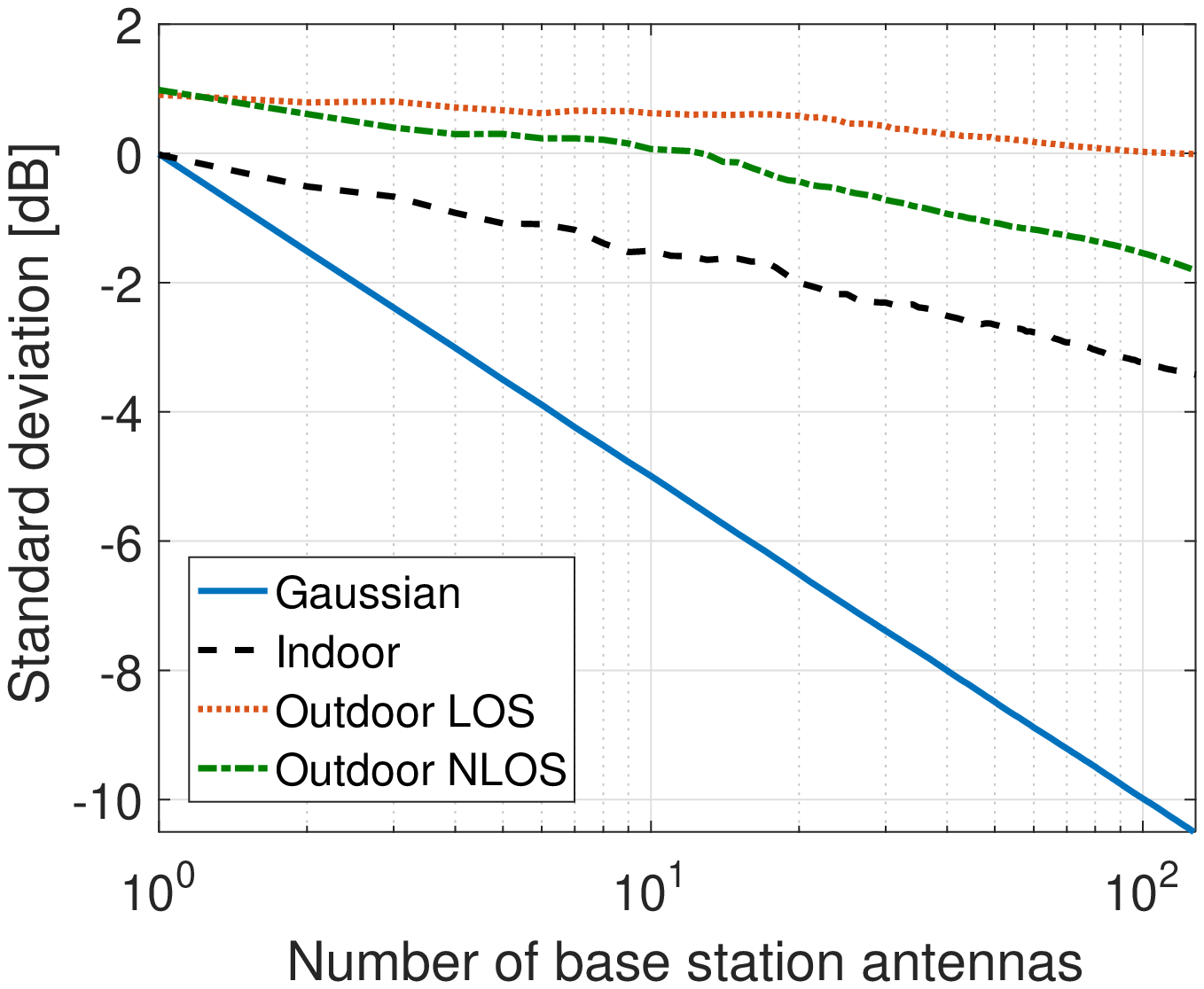}}
	\caption{The users with the least channel hardening}
	\label{fig:in_vs_out_worst}
\end{subfigure}
\caption{Standard deviation of channel gain as a function of number of base station antennas for the three different scenarios with the cylindrical array, and the complex Gaussian channel as reference.}
\label{fig:scenarios}
\end{figure*}

To investigate differences and similarities between scenarios, the channel hardening is also analysed for measurements from the outdoor scenario, both in LOS and NLOS, as depicted in Fig.~\ref{fig:location2}. For all three scenarios, the users with the most and least channel hardening are shown in Fig.~\ref{fig:in_vs_out_best} and Fig.~\ref{fig:in_vs_out_worst}, respectively. Note that these are not necessarily the users with the smallest and largest standard deviation for a specific subset of antennas, such as for $M=128$.
In Fig.~\ref{fig:in_vs_out_best} it can be seen that the user with the most channel hardening is in the outdoor NLOS scenario. However, the channel from the indoor scenario could already be interpreted as being 'pre-hardened', as it for each subset of antennas already has a smaller standard deviation of channel gain relative to its mean. The users with the least channel hardening, and also the largest standard deviation of channel gain for $M=128$, are in general found in the outdoor LOS scenario.
Recalling that the channels for the different users are normalized, it should be noted that even though users in the outdoor LOS scenario have the least channel hardening it does not necessarily imply that it is a bad channel since the un-normalized channel could still have a considerable high received channel gain and that a strong LOS in general gives good communication performance.
A comparison of the mean and maximum un-normalized channel gain in the three scenarios is found in Table~\ref{table:gain}. It can be observed that it is indeed the case that both the mean and the maximum channel gain, as per antenna, are as expected higher in the outdoor LOS scenario than in the NLOS scenario, although lower than in the indoor scenario. However, the distances are different and thereby also the pathloss.

\begin{table}[b]
    \normalsize
    \centering
    \begin{tabular}{c | c | c | c}
         & Scenario & Mean & Maximum \\
         \hline
         Fig.~\ref{fig:in_vs_out_best} & Indoor & -71 & -66 \\ 
         & Outdoor LOS & -100 & -93 \\ 
         & Outdoor NLOS & -107 & -104  \\ 
         \hline
          Fig.~\ref{fig:in_vs_out_worst} & Indoor & -72 & -66\\ 
         & Outdoor LOS & -88 &  -81 \\ 
         & Outdoor NLOS & -105 & -99 
    \end{tabular}
    \caption{Comparison of the antennas with the mean and maximum average un-normalized channel gain in the different scenarios.}
    \label{table:gain}
\end{table}

\begin{figure*}[t]
\hspace{-0.5in}
\begin{subfigure}[b]{0.4\textwidth}
    \centering
    \hspace{0.4in}
    \psfrag{1}[][]{\Large \hspace{-0.15in} 1}
    \psfrag{2}[][]{\Large 2}
    \psfrag{0}[][]{\Large 0}
    \psfrag{3}[][]{\Large 3}
    \psfrag{4}[][]{\Large 4}
    \psfrag{5}[][]{\Large 5}
    \psfrag{6}[][]{\Large 6}
    \psfrag{7}[][]{\Large 7}
    \psfrag{8}[][]{\Large 8}
    \psfrag{9}[][]{\Large 9}
    \psfrag{0.1}[][]{\Large \hspace{-0.15in} 0.1}
    \psfrag{0.2}[][]{\Large \hspace{-0.15in} 0.2}
    \psfrag{0.3}[][]{\Large \hspace{-0.15in} 0.3}
    \psfrag{0.4}[][]{\Large \hspace{-0.15in} 0.4}
    \psfrag{0.5}[][]{\Large \hspace{-0.15in} 0.5}
    \psfrag{0.6}[][]{\Large \hspace{-0.15in} 0.6}
    \psfrag{0.7}[][]{\Large \hspace{-0.15in} 0.7}
    \psfrag{0.8}[][]{\Large \hspace{-0.15in} 0.8}
    \psfrag{0.9}[][]{\Large \hspace{-0.15in} 0.9}
    \psfrag{F(x)}[][]{\Large F(x)}
    \psfrag{x}[][]{\Large Normalized channel gain [dB]}
    \psfrag{Empirical CDF}[][]{}
    \resizebox{2.5in}{!}{\includegraphics[width=4in]{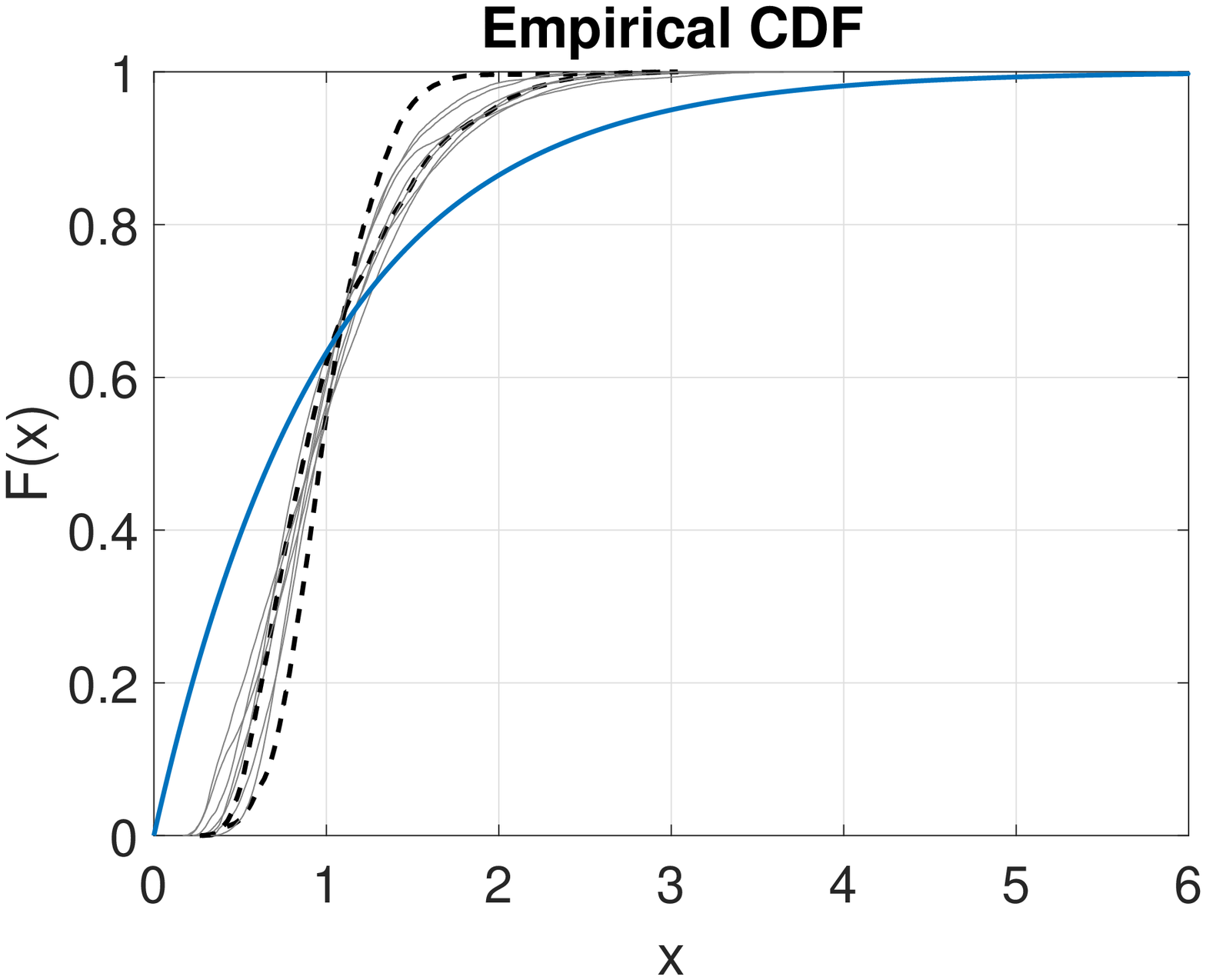}}
	\caption{Indoor}
	\label{fig:cdfs}
\end{subfigure}
\hspace{-0.5in}
\begin{subfigure}[b]{0.4\textwidth}
    \centering
    \psfrag{1}[][]{\Large \hspace{-0.15in} 1}
    \psfrag{2}[][]{\Large 2}
    \psfrag{0}[][]{\Large 0}
    \psfrag{3}[][]{\Large 3}
    \psfrag{4}[][]{\Large 4}
    \psfrag{5}[][]{\Large 5}
    \psfrag{6}[][]{\Large 6}
    \psfrag{7}[][]{\Large 7}
    \psfrag{8}[][]{\Large 8}
    \psfrag{9}[][]{\Large 9}
    \psfrag{0.1}[][]{\Large \hspace{-0.15in} 0.1}
    \psfrag{0.2}[][]{\Large \hspace{-0.15in} 0.2}
    \psfrag{0.3}[][]{\Large \hspace{-0.15in} 0.3}
    \psfrag{0.4}[][]{\Large \hspace{-0.15in} 0.4}
    \psfrag{0.5}[][]{\Large \hspace{-0.15in} 0.5}
    \psfrag{0.6}[][]{\Large \hspace{-0.15in} 0.6}
    \psfrag{0.7}[][]{\Large \hspace{-0.15in} 0.7}
    \psfrag{0.8}[][]{\Large \hspace{-0.15in} 0.8}
    \psfrag{0.9}[][]{\Large \hspace{-0.15in} 0.9}
    \psfrag{F(x)}[][]{\Large F(x)}
    \psfrag{x}[][]{\Large Normalized channel gain [dB]}
    \psfrag{Empirical CDF}[][]{}
    \resizebox{2.5in}{!}{\includegraphics[width=4in]{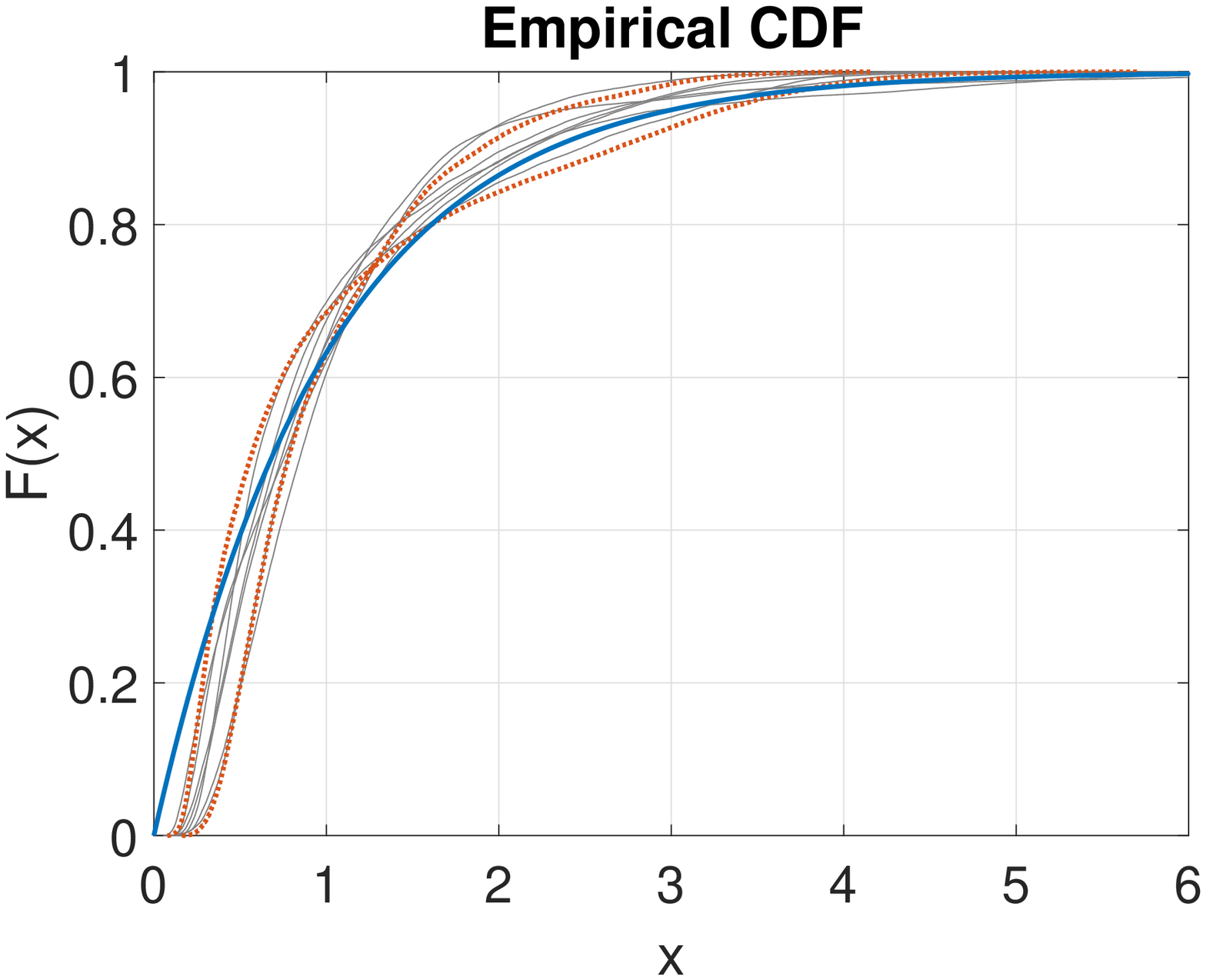}}
	\caption{Outdoor LOS}
	\label{fig:cdfs}
\end{subfigure}
\hspace{-0.5in}
\begin{subfigure}[b]{0.4\textwidth}
    \centering
    \psfrag{1}[][]{\Large \hspace{-0.15in} 1}
    \psfrag{2}[][]{\Large 2}
    \psfrag{0}[][]{\Large 0}
    \psfrag{3}[][]{\Large 3}
    \psfrag{4}[][]{\Large 4}
    \psfrag{5}[][]{\Large 5}
    \psfrag{6}[][]{\Large 6}
    \psfrag{7}[][]{\Large 7}
    \psfrag{8}[][]{\Large 8}
    \psfrag{9}[][]{\Large 9}
    \psfrag{0.1}[][]{\Large \hspace{-0.15in} 0.1}
    \psfrag{0.2}[][]{\Large \hspace{-0.15in} 0.2}
    \psfrag{0.3}[][]{\Large \hspace{-0.15in} 0.3}
    \psfrag{0.4}[][]{\Large \hspace{-0.15in} 0.4}
    \psfrag{0.5}[][]{\Large \hspace{-0.15in} 0.5}
    \psfrag{0.6}[][]{\Large \hspace{-0.15in} 0.6}
    \psfrag{0.7}[][]{\Large \hspace{-0.15in} 0.7}
    \psfrag{0.8}[][]{\Large \hspace{-0.15in} 0.8}
    \psfrag{0.9}[][]{\Large \hspace{-0.15in} 0.9}
    \psfrag{F(x)}[][]{\Large F(x)}
    \psfrag{x}[][]{\Large Normalized channel gain [dB]}
    \psfrag{Empirical CDF}[][]{}
    \resizebox{2.5in}{!}{\includegraphics[width=4in]{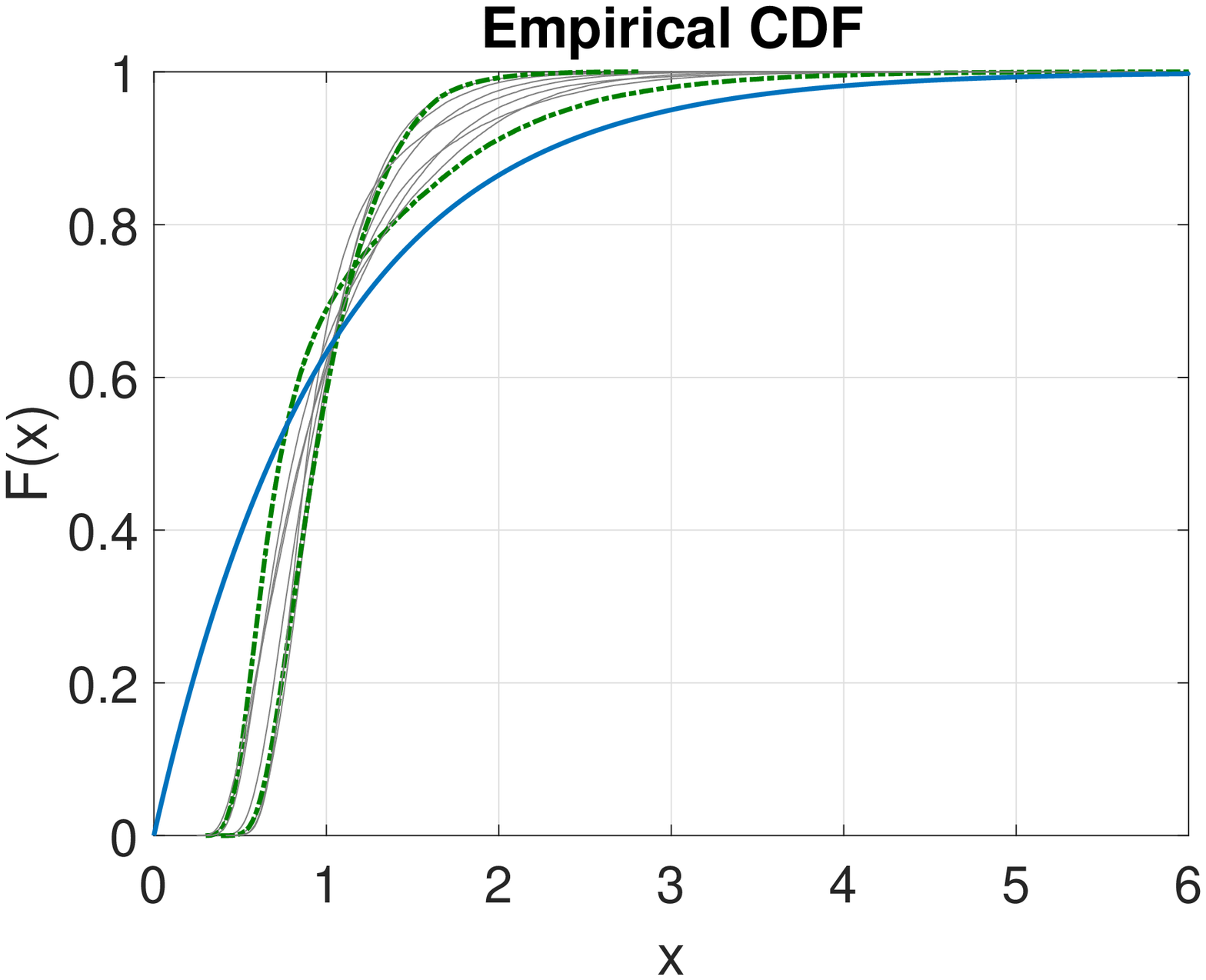}}
	\caption{Outdoor NLOS}
\end{subfigure}
	\caption{Empirical CDFs of the normalized channel gain when using all 128 antennas in the cylindrical array, for all scenarios and users. The users marked in (a) black dashed, (b) orange dotted and (c) green dash-dotted lines have the least and the most channel hardening, and correspond to the users shown in Fig.~\ref{fig:scenarios}. To serve as reference, the exponential distribution ($\lambda=1$) is in blue full line.}
	\label{fig:cdfs}
\end{figure*}

\begin{table}[b]
    \normalsize
    \centering
    \begin{tabular}{c | c | c | c }
         & Scenario & Mean ratio & Std ratio \\
         \hline
         Fig.~\ref{fig:in_vs_out_best} & Indoor & -0.7 & 8.0 \\
         & Outdoor LOS & -0.2 & 7.8 \\
         & Outdoor NLOS & -0.3 & 7.8 \\ 
         \hline
          Fig.~\ref{fig:in_vs_out_worst} & Indoor & 1.5 & 7.9\\
         & Outdoor LOS & 1.4 & 7.9\\ 
         & Outdoor NLOS & 1.1 & 7.9 
    \end{tabular}
    \caption{Mean and standard deviation (std) of the ratio between the channel gains of the vertical and horizontal polarization for the users in Fig.~\ref{fig:scenarios}.}
    \label{table:pol_ratio}
\end{table}

Fig.~\ref{fig:cdfs} shows the empirical CDFs of their channel gains when $M=128$ for all scenarios and users, providing further insights about the behaviour of the curves in the different scenarios. The marked users experience the least and most channel hardening, also shown in Fig.~\ref{fig:in_vs_out_best} and \ref{fig:in_vs_out_worst}, respectively. Evidently, the steeper the CDF is, the more channel hardening. Another observation is that if there are strong outliers, the standard deviation level is increased; this is the case for the outdoor scenarios. As a reference, the exponential distribution with $\lambda=1$ is also provided, corresponding to the channel gain, i.e. $|h|^2$, with a mean of $1$ for the one-antenna case where the channel is complex Gaussian distributed; this means that there is no channel hardening. Given this, the observation is that although the outdoor LOS scenario is to a large extent similar to the reference curve for higher channel gains, the risk of lower gains, affecting the probability of outage, is  reduced when combining the many antennas. As a final remark on the comparison between different scenarios, a quantification of the average channel hardening results in $3.9$~dB for the indoor scenario, $1.7$~dB for the outdoor LOS scenario and $4.3$~dB for the outdoor NLOS scenario.

\subsection{Influence of polarization}

After exploring the influence of LOS and NLOS on the experienced channel hardening, further investigations included a revisit to take a closer look into the polarization aspect. In Table~\ref{table:pol_ratio}, the mean and standard deviation of the ratio between the channel gains of the vertical and horizontal polarization, as per snapshot, frequency point and antenna element, are shown for the users in Fig.~\ref{fig:scenarios}. Although, they all seem to have similar standard deviation, the mean is in general closer to zero for the users with the most channel hardening; this is likely contributing to a more even distribution of the channel gain.

\begin{figure}[t]
	\centering
    \psfrag{2}[][]{\Large 2}
    \psfrag{0}[][]{\Large 0}
    \psfrag{-2}[][]{\Large -2}
    \psfrag{-4}[][]{\Large -4}
    \psfrag{-6}[][]{\Large -6}
    \psfrag{-8}[][]{\Large -8}
    \psfrag{-10}[][]{\Large -10}
    \psfrag{10}[][]{\Large 10 \hspace{0.025in}}
    \psfrag{1}[][]{\Large 1}
    \psfrag{Number of base station antennas}[][][1.1]{\Large Number of base station antennas}
    \psfrag{Standard deviation [dB]}[][][1.1]{\Large Standard deviation [dB]}
    \psfrag{Gaussian}[][]{\normalsize Gaussian}
    \psfrag{Cylindrical user 1}[][]{\normalsize Cylindrical user 1}
    \psfrag{Cylindrical user 5}[][]{\normalsize Cylindrical user 5}
    \psfrag{Planar user 1}[][]{\normalsize Planar user 1}
    \psfrag{Planar user 5}[][]{\normalsize Planar user 5}
    \resizebox{3in}{!}{\includegraphics[width=4in]{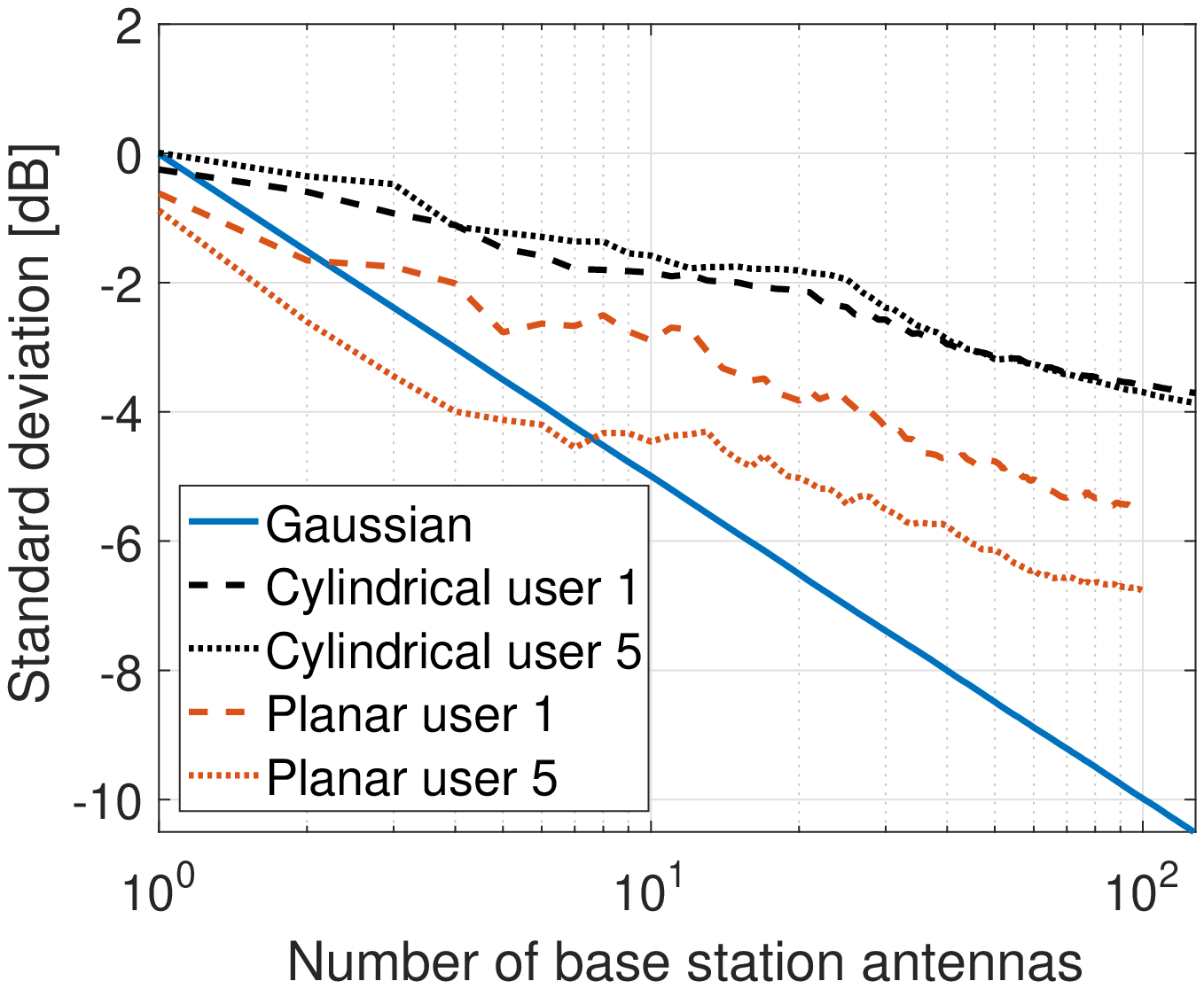}}
	\caption{Standard deviation of channel gain as a function of number of base station antennas for user~1 and~5, and the complex Gaussian channel as reference. The scenario is the indoor scenario where the users are surrounded by a crowd with the cylindrical and the planar array, respectively.}
	\label{fig:cyl_vs_plan}
\end{figure}

\subsection{Interaction between environment and antenna array}

\begin{figure}[t]
	\centering
    \psfrag{1}[][]{\Large \hspace{-0.15in} 1}
    \psfrag{2}[][]{\Large 2}
    \psfrag{1.5}[][]{\Large 1.5}
    \psfrag{2.5}[][]{\Large 2.5}
    \psfrag{0}[][]{\Large 0}
    \psfrag{3}[][]{\Large 3}
    \psfrag{4}[][]{\Large 4}
    \psfrag{5}[][]{\Large 5}
    \psfrag{6}[][]{\Large 6}
    \psfrag{7}[][]{\Large 7}
    \psfrag{8}[][]{\Large 8}
    \psfrag{9}[][]{\Large 9}
    \psfrag{0.1}[][]{\Large \hspace{-0.15in} 0.1}
    \psfrag{0.2}[][]{\Large \hspace{-0.15in} 0.2}
    \psfrag{0.3}[][]{\Large \hspace{-0.15in} 0.3}
    \psfrag{0.4}[][]{\Large \hspace{-0.15in} 0.4}
    \psfrag{0.5}[][]{\Large \hspace{-0.15in} 0.5}
    \psfrag{0.6}[][]{\Large \hspace{-0.15in} 0.6}
    \psfrag{0.7}[][]{\Large \hspace{-0.15in} 0.7}
    \psfrag{0.8}[][]{\Large \hspace{-0.15in} 0.8}
    \psfrag{0.9}[][]{\Large \hspace{-0.15in} 0.9}
    \psfrag{F(x)}[][]{\Large F(x)}
    \psfrag{x}[][]{\Large Normalized channel gain [dB]}
    \psfrag{Empirical CDF}[][]{}
    \psfrag{Exponential}[][]{\normalsize Exponential}
    \psfrag{Cylindrical user 1}[][]{\normalsize Cylindrical user 1}
    \psfrag{Cylindrical user 5}[][]{\normalsize Cylindrical user 5}
    \psfrag{Planar user 1}[][]{\normalsize Planar user 1}
    \psfrag{Planar user 5}[][]{\normalsize Planar user 5}
    \resizebox{3in}{!}{\includegraphics[width=4in]{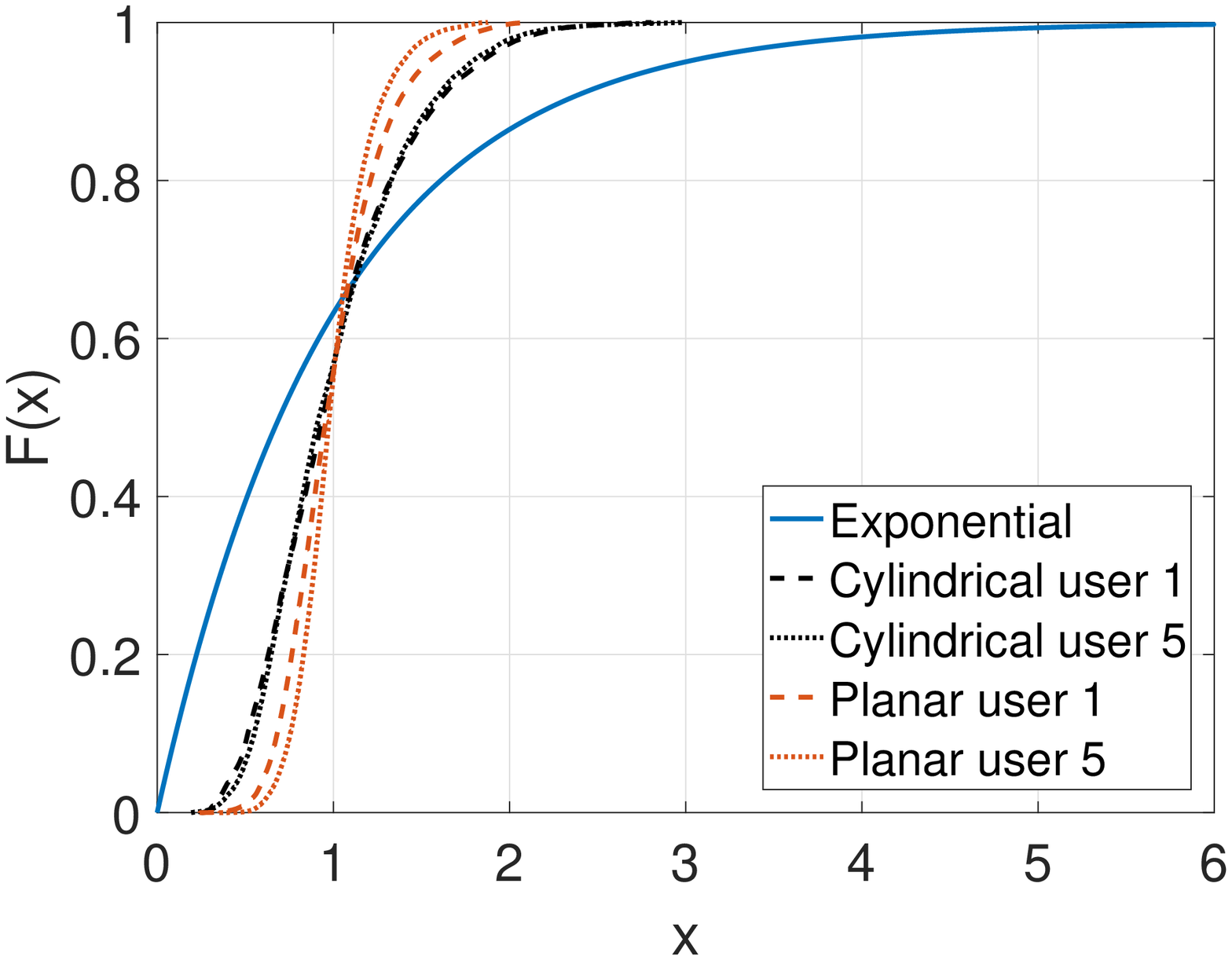}}
	\caption{Empirical CDFs of the normalized channel gain when using all 100 or 128 antennas for the two arrays and users in Fig.~\ref{fig:cyl_vs_plan}. and the exponential distribution ($\lambda=1$) as reference.}
	\label{fig:cdf_arrays}
\end{figure}

The last aspect we want to further investigate is the interaction between the environment and the array. Therefore, measurements were also performed in the indoor scenario, as previously has been described (see Fig.~\ref{fig:location}), with the planar array shown in Fig.~\ref{fig:lumami} for user~1 and 5. The resulting channel hardening curves are shown in Fig.~\ref{fig:cyl_vs_plan}. For these measurements the users were surrounded with a crowd, such that the square in Fig.~\ref{fig:location} was almost filled with people. The results in Fig.~\ref{fig:cyl_vs_plan} show that significantly smaller standard deviations of channel gain, i.e. down to -6.8~dB, are achieved for both users when using the planar array in comparison to using the cylindrical array. This demonstrates that for a typical deployment geometry, the planar array can better exploit the diversity in the environment as more of the antennas are effective and the distribution of channel gain over the array is more even; the latter can be seen in Fig.~\ref{fig:cdf_arrays} where the slopes of the empirical CDFs of the normalized channel gain for the scenario with the planar array are steeper.

%% file: cost2100.tex
This section provides a short description of the COST 2100 channel model, the simulations performed with the model, which are later used for comparison, and an example result from the simulations. Initial answers to the questions if and how well the COST 2100 channel model with its model parameters in general can capture the channel hardening effect are also given. This is further elaborated on in the comparison in Section~VI.

The COST 2100 channel model \cite{cost2100}, including the massive MIMO extension described in \cite{cost_pepe}, is a GSCM that can stochastically describe massive MIMO channels over time, frequency and space in a spatially consistent manner. The goal with the extension is to capture channel characteristics that become more prominent in massive MIMO channels. These extensions are:

\begin{itemize}
    \item 3D propagation,
    \item the possibility for different parts of a physically large array to experience different clusters, and
    \item a gain-function regulating the gain for individual multipath components, which is of particular importance to closely-spaced users.
\end{itemize}

For the evaluation in this paper we use the indoor scenario with the cylindrical array and closely-spaced users at 2.6~GHz. This scenario has been parameterized based on the measurement data used in previous experimental analysis and the complete model can be found at \cite{cost_model}. More specifically, the chosen settings when running the simulations are outlined in Table~\ref{table:settings}.

\begin{table}[b]
    \normalsize
    \centering
    \begin{tabular}{c | c}
         & Setting \\
         \hline
         Network & 'Indoor\_CloselySpacedUser\_2\_6GHz'\\
         Link & 'Multiple' \\
         Antenna & 'MIMO\_Cyl\_patch' \\
         Band & 'Wideband' 
    \end{tabular}
    \caption{Settings chosen for the simulations with the COST 2100 channel model.}
    \label{table:settings}
\end{table}

In the simulations, 300~snapshots were generated at a rate of 50~snapshots per second with a user velocity of 0.25~m/s; this means that samples are taken over a distance of 1.5~m in total. 129 points in frequency are computed. The simulated propagation environment is combined with a synthetic antenna pattern of a cylindrical array with 128 antennas at the base station side. At the user side, the multipath components obtained from the simulated environment are combined with either an antenna pattern with user effect or an omni-directional antenna, for comparison. A description of the antenna pattern with user effect can be found in~\cite{ant_pattern}.

\begin{figure}[t]
	\centering
    \psfrag{2}[][]{\Large 2}
    \psfrag{0}[][]{\Large 0}
    \psfrag{-2}[][]{\Large -2}
    \psfrag{-4}[][]{\Large -4}
    \psfrag{-6}[][]{\Large -6}
    \psfrag{-8}[][]{\Large -8}
    \psfrag{-10}[][]{\Large -10}
    \psfrag{10}[][]{\Large 10 \hspace{0.025in}}
    \psfrag{1}[][]{\Large 1}
    \psfrag{Number of base station antennas}[][][1.1]{\Large Number of base station antennas}
    \psfrag{Standard deviation [dB]}[][][1.1]{\Large Standard deviation [dB]}
    \psfrag{Gaussian}[][]{\normalsize Gaussian}
    \psfrag{COST omni}[][]{\normalsize COST omni}
    \psfrag{COST meas}[][]{\normalsize COST meas}
    \psfrag{Meas best}[][]{\normalsize Meas best}
    \psfrag{Meas worst}[][]{\normalsize Meas worst}
    \resizebox{3in}{!}{\includegraphics[width=4in]{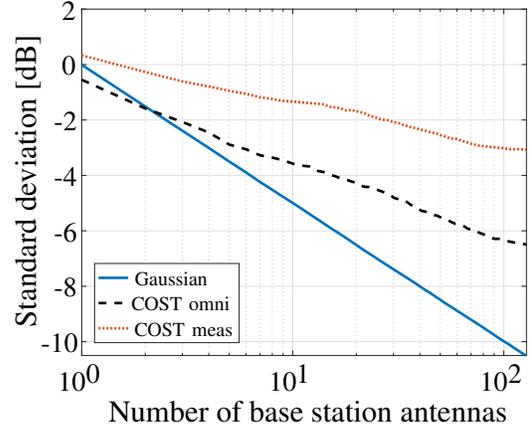}}
	\caption{Standard deviation of channel gain as a function of number of base station antennas with an omni-directional antenna (COST omni) or an antenna pattern with user effect (COST meas) in the COST 2100 channel model for a typical user, and the complex Gaussian channel as a reference.}
	\label{fig:omni_vs_meas_ue1}
\end{figure}

Each user antenna starts with a random initial orientation uniformly generated between $[-\pi \; \pi)$, and during the simulation each user antenna undergoes a randomly generated rotation between $[-\pi \; \pi)$ in total in the azimuth plane. For the obtained transfer function, the channels are normalized as in (\ref{eq:norm}) and the standard deviation of channel gain is computed as in (\ref{eq:std}), in the same manner as for the measurement data. Running the simulation ten times, and averaging the resulting standard deviation of channel gain over these ten simulations, gives the result in Fig.~\ref{fig:omni_vs_meas_ue1} for a typical user.

 As an example result from the simulations, Fig.~\ref{fig:omni_vs_meas_ue1} shows the resulting standard deviation both when using the omni-directional and measured antenna pattern with user effect, for the same propagation environment. The complex Gaussian channel is also shown as a reference.
 The difference of the starting point of the two curves with different antenna patterns is almost 0.9~dB while at the end point, the difference is about 3.4~dB. 
 The result here is only for one specific user antenna pattern, and changing this pattern would affect the channel hardening. However, it shows that there is both a limitation on the channel hardening that is imposed by the simulated environment, in line with the presented theory in Section~III, but also that the antenna pattern of the user can reduce the experienced channel hardening as much as to half of the channel hardening experienced with an omni-directional antenna, on a linear scale. In general, any user equipment with a non omni-directional antenna pattern will likely degrade the experienced channel hardening as not all scatterers will be effective.

%% file: comparison.tex
The last part of this study is a joint comparison of the three levels at which channel hardening has been analyzed: theory, measurements and simulations. This provides a more thorough analysis of channel hardening and synthesizes the three points of view. We discuss aspects which are relevant when designing reliable massive MIMO systems and highlight channel and antenna characteristics important for channel hardening.

\begin{figure*}[t]
\centering
    \begin{subfigure}[b]{0.4\textwidth}
    \psfrag{0}[][]{\Large 0}
    \psfrag{1}[][]{\Large 1}
    \psfrag{2}[][]{\Large 2}
    \psfrag{-2}[][]{\Large -2}
    \psfrag{-4}[][]{\Large -4}
    \psfrag{-6}[][]{\Large -6}
    \psfrag{-8}[][]{\Large -8}
    \psfrag{-10}[][]{\Large -10}
    \psfrag{10}[][]{\Large 10 \hspace{0.025in}}
    \psfrag{20}[][]{\Large 20}
    \psfrag{40}[][]{\Large 40}
    \psfrag{60}[][]{\Large 60}
    \psfrag{80}[][]{\Large 80}
    \psfrag{100}[][]{\Large 100}
    \psfrag{120}[][]{\Large 120}
    \psfrag{Number of base station antennas}[][][1.1]{\Large Number of base station antennas}
    \psfrag{Standard deviation [dB]}[][][1.1]{\Large Standard deviation [dB]}
    \psfrag{Gaussian}[][]{\normalsize Gaussian}
    \psfrag{Measurement}[][]{\normalsize Measurement}
    \psfrag{COST omni}[][]{\normalsize \hspace{0.05in} COST omni}
    \psfrag{COST meas}[][]{\normalsize \hspace{0.05in} COST meas}
    \psfrag{P=13}[][]{\normalsize \hspace{0.05in} $P=13$}
    \psfrag{P=29}[][]{\normalsize \hspace{0.05in} $P=29$}
    \psfrag{P=5}[][]{\normalsize \hspace{0.05in} $P=5$}
    \psfrag{P=22}[][]{\normalsize \hspace{0.05in} $P=22$}
    \resizebox{3in}{!}{\includegraphics[width=4in]{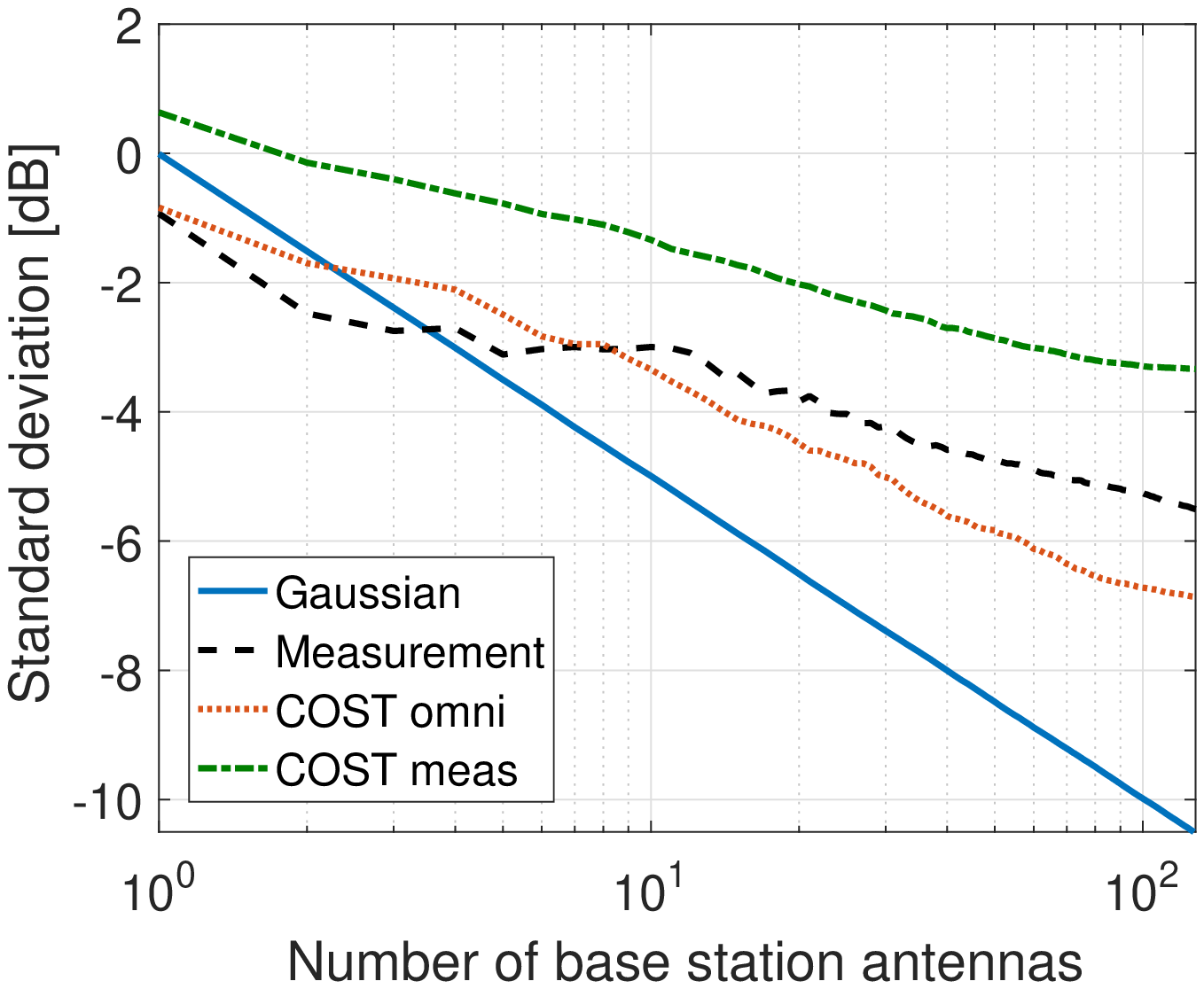}}
	\caption{The users with the most channel hardening}
	\label{fig:comp_best}
	\end{subfigure}
	\begin{subfigure}[b]{0.4\textwidth}
    \psfrag{0}[][]{\Large 0}
    \psfrag{1}[][]{\Large 1}
    \psfrag{2}[][]{\Large 2}
    \psfrag{-2}[][]{\Large -2}
    \psfrag{-4}[][]{\Large -4}
    \psfrag{-6}[][]{\Large -6}
    \psfrag{-8}[][]{\Large -8}
    \psfrag{-10}[][]{\Large -10}
    \psfrag{10}[][]{\Large 10 \hspace{0.025in}}
    \psfrag{20}[][]{\Large 20}
    \psfrag{40}[][]{\Large 40}
    \psfrag{60}[][]{\Large 60}
    \psfrag{80}[][]{\Large 80}
    \psfrag{100}[][]{\Large 100}
    \psfrag{120}[][]{\Large 120}
    \psfrag{Number of base station antennas}[][][1.1]{\Large Number of base station antennas}
    \psfrag{Standard deviation [dB]}[][][1.1]{\Large Standard deviation [dB]}
    \psfrag{Gaussian}[][]{\normalsize Gaussian}
    \psfrag{Measurement}[][]{\normalsize Measurement}
    \psfrag{COST omni}[][]{\normalsize \hspace{0.05in} COST omni}
    \psfrag{COST meas}[][]{\normalsize \hspace{0.05in} COST meas}
    \psfrag{P=4}[][]{\normalsize \hspace{0.05in} $P=4$}
    \psfrag{P=11}[][]{\normalsize \hspace{0.05in} $P=11$}
    \psfrag{P=3}[][]{\normalsize \hspace{0.05in} $P=3$}
    \psfrag{P = 22}[][]{\normalsize \hspace{0.05in} $P=22$}
    \resizebox{3in}{!}{\includegraphics[width=4in]{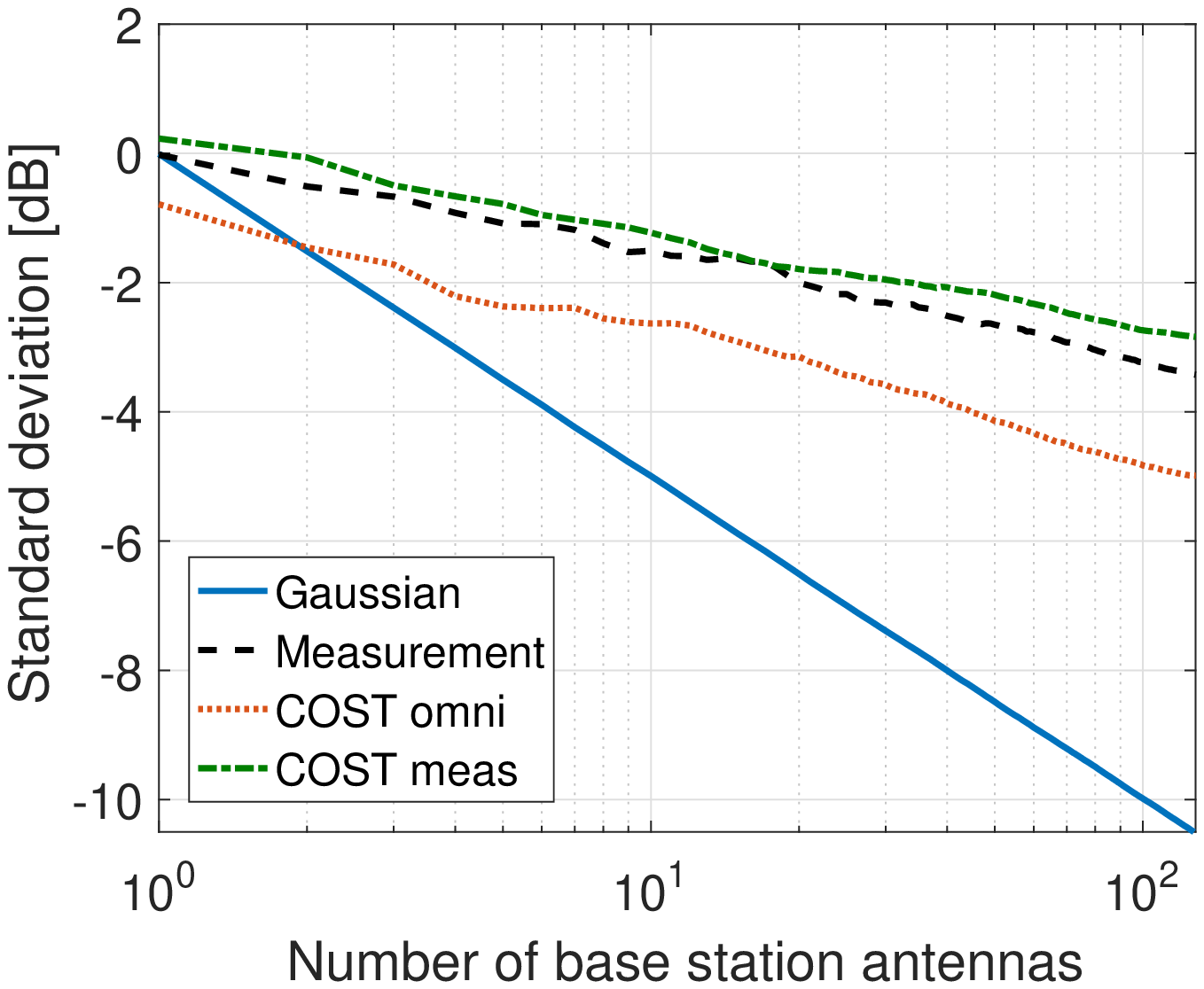}}
	\caption{The users with the least channel hardening}
	\label{fig:comp_worst}
	\end{subfigure}
\caption{Standard deviation of channel gain as a function of number of base station antennas for one user from (i) the measurements, (ii) the simulation with an omni-directional antenna and (iii) the simulation with an antenna pattern with user effect in the indoor scenario with the cylindrical array, and the complex Gaussian channel as reference. }
\label{fig:comparison}
\end{figure*}

\begin{figure}[t]
	\centering
    \psfrag{2}[][]{\Large 2}
    \psfrag{0}[][]{\Large 0}
    \psfrag{-0.5}[][]{\Large -0.5}
    \psfrag{-1}[][]{\Large -1}
    \psfrag{-1.5}[][]{\Large -1.5}
    \psfrag{-2}[][]{\Large -2}
    \psfrag{-10}[][]{\Large -10}
    \psfrag{10}[][]{\Large 10 \hspace{0.025in}}
    \psfrag{1}[][]{\Large 1}
    \psfrag{X}[][][1.1]{\Large Number of base station antennas}
    \psfrag{Y}[][][1.1]{\Large $\Delta$ of standard deviation [dB]}
    \psfrag{Gaussian}[][]{\normalsize Gaussian}
    \psfrag{COST omni}[][]{\normalsize COST omni}
    \psfrag{COST meas}[][]{\normalsize COST meas}
    \psfrag{Measurement}[][]{\normalsize Measurement}
    \psfrag{Meas worst}[][]{\normalsize Meas worst}
    \resizebox{3in}{!}{\includegraphics[width=4in]{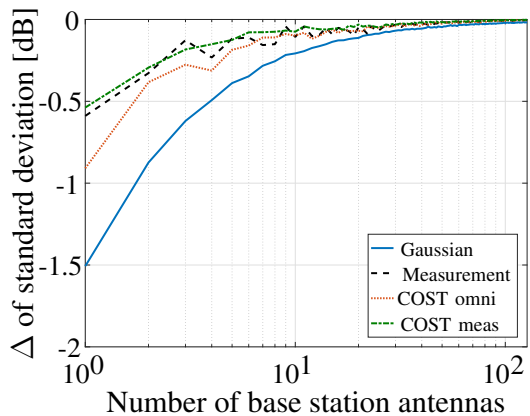}}
	\caption{Decrease ($\Delta$) of standard deviation of channel gain as a function of the increasing number of base station antennas for the average user in (i) the measurements, (ii) the simulation with an omni-directional antenna and (iii) the simulation with an antenna pattern with user effect, and the complex Gaussian channel as reference.}
	\label{fig:slope}
\end{figure}

In Fig.~\ref{fig:comparison}, a comparison between theory, measurements and simulation results from the COST 2100 channel model is shown. The users shown from the measurements and simulations are the users with the most channel hardening (Fig.~\ref{fig:comp_best}) and the least channel hardening, (Fig.~\ref{fig:comp_worst}) among the nine users when using all 128~base station antennas. As reference, the complex Gaussian channel is included.

Quantifying the general behaviour of the curves from the simulations, Fig.~\ref{fig:slope} shows the decrease ($\Delta$) of standard deviation when increasing the number of antennas, averaged over all users. Here, it can be seen that on average, the slopes resulting from the measurements and the COST simulations with measured antenna pattern have very similar behaviors and that the model in general shows good correspondence. Meanwhile, the COST simulations with an omni-directional antenna have in general a larger slope, although not as high as in the complex Gaussian case due to the limitations in the environment. Considering the average channel hardening in the three cases, the measurements give a channel hardening of 3.9~dB, while the COST simulations on average give, for the measured and omni-directional antenna, a channel hardening of 3.5~dB and 5.3~dB, respectively. Looking at the average starting points for the three cases, the measurements start at 0.3~dB while the COST simulations with the measured antenna pattern start at 0.6~dB and the omni-directional antenna at \\-0.8~dB.
Below we summarize the results and discussions from three important aspects.

\textbf{\textit{The complex Gaussian channel model is overoptimistic:}} Starting the discussion from a theoretical point of view, the complex Gaussian channel, which is only limited by the number of antennas, gives a channel hardening of 10.5~dB when having 128~base station antennas. This channel has often been assumed in theoretical studies of massive MIMO and is shown in this study to be more optimistic than real measured massive MIMO channels, which in general are spatially correlated. In reality the environment does generally not create as rich scattering as assumed in the complex Gaussian channel. As a result, channel responses are rather spatially correlated, and thus the environment also imposes a limitation on the achievable channel hardening; this fact is captured by (\ref{eq:cv}). Moreover, the interactions between the array and the environment can result in quite different received power levels at the antenna elements.

\textbf{\textit{The channel gain distribution over the array is essential:}} Repeating the assumptions made in \cite{chhard_phy_model} in order to get to (\ref{eq:cv}), it was assumed that the channel coefficients for all physical paths are $\mathcal{CN}(0,1)$ and that the steering vectors of the physical paths are distributed uniformly over the unit sphere. If the mean $\mu$ and the variance $\sigma^2$ would deviate from this assumption, then the large-scale fading term in (\ref{eq:cv}) would be bounded by $\frac{1}{P}(\frac{\sigma}{\mu})^2$, see \cite{chhard_phy_model}. With a large variance, relative to the mean, the standard deviation curves would move up. The curves would move down if the variance is small relative to the mean. What can be seen in e.g. Figs~8-10 is indeed that when having a higher mean, the standard deviation curves in the indoor scenario move down; this is likely due to having a strong LOS component. Having a large imbalance in the channel gain over the array causes variations and can hence move the standard deviation curves up; causes of these variations can be due to having some antennas experiencing a strong LOS while others are not. This imbalance can also occur between the polarization modes. These aspects could as well change over time and thereby much of the remaining variations of channel gain come from the time domain.

\textbf{\textit{The user movements and antenna patterns are uncertain:} }
The large variation of channel gain in the time domain can be hard to predict as there is always an uncertainty in the user movements and as a result of this, in the interaction between the environment and the antenna patterns of the user equipment. It is clear that the user antenna pattern and behaviour can be a limiting factor for the channel hardening. This can be seen in Fig.~\ref{fig:comparison} as the simulated curves with the antenna pattern with user effect are close to the user with the least channel hardening in the measurements in Fig.~\ref{fig:comp_worst}, while a user with an omni-directional antenna experiences more channel hardening. Assuming a non-beneficial scenario from the user aspect, then this type of simulation, where the user equipment is rotating, could be used while in reality it could happen that conditions are more favourable and thus a smaller standard deviation could be achieved. For example, the user from the measurements with the most channel hardening in Fig.~\ref{fig:comp_best}, experiences a similar channel hardening as the simulations with the omni-directional antenna in Fig.~\ref{fig:comp_worst}. Overall, the most significant reasons for the differences between the measurements and the simulations in Fig.~\ref{fig:comparison} are the different antenna patterns and movements of the users; to further align the results, then these parameters should be the same. Further on, since the COST 2100 channel model is a stochastic channel model, comparisons with particular measurements are difficult. However, on average the channel hardening curves resulting from the model show very similar behavior as for the measurements, as seen in Fig.~\ref{fig:slope}.

%% file: conclusion.tex
From our experiments and analysis, we can conclude that there is a clear channel hardening effect in massive MIMO, which flattens out the channel fading in frequency and time as the number of antennas increases. However, the complex i.i.d. Gaussian channel model, which is commonly assumed in theoretical studies, is shown to be overoptimistic and does not provide an accurate model for channel hardening in massive MIMO. This is due to the fact that real environments do not provide as rich scattering as assumed in the model but are rather spatially correlated, as recently acknowledged in literature. Moreover, interactions between the array and the environment will cause variations of the received power levels between the different antenna elements. Therefore, by investigating the mechanisms that build up the distribution of channel gain over the array, the channel hardening effect can be better understood. Power imbalances can be due to having a set of antennas where some have a stronger LOS and some not; this imbalance could also be between the polarization modes. As these imbalances change over time, there will still be variations of the channel gain that are originating from the time domain. These changes are hard to predict as there is always an uncertainty of user movements and antenna pattern.

We have also shown that the COST 2100 model with its massive MIMO extension statistically can capture the channel hardening effect well, given appropriate propagation parameters and a realistic antenna pattern and user behavior.
As a final remark, in order to maximize the experienced channel hardening, one needs to find strategies to cope with the power imbalances over the array and design the massive MIMO system such that it can best exploit the diversity of the considered environment. However, channel hardening is only one part of the complex story of achieving reliable communication and in future work other aspects, such as received power levels and coverage, should also be considered.